\documentstyle[floats,aps,epsf,amsfonts]{revtex}
\begin{document}

\title{Regularization parameters for the self force in Schwarzschild
spacetime:\\ I. scalar case}
\author{Leor Barack$^1$ and Amos Ori$^2$}
\address{
$^1$Albert-Einstein-Institut, Max-Planck-Institut f{\"u}r
Gravitationsphysik, Am M\"uhlenberg 1, D-14476 Golm, Germany.
$^2$Department of Physics, Technion---Israel Institute of Technology,
Haifa, 32000, Israel.\\
}
\date{\today}
\maketitle


\begin{abstract}

We derive the explicit values of all regularization parameters (RP)
for a scalar particle in an arbitrary geodesic orbit around a Schwarzschild
black hole. These RP are required within the previously introduced mode-sum
method, for calculating the local self-force acting on the particle.
In this method one first calculates the (finite) contribution to the
self-force due to each individual multipole mode of the particle's
field, and then applies a certain regularization procedure to the mode sum,
involving the RP.  The explicit values of the RP were presented in a recent
Letter [Phys.\ Rev.\ Lett.\ {\bf 88}, 091101 (2002)]. Here we give full details
of the RP derivation in the scalar case. The calculation of the RP in the
electromagnetic and gravitational cases will be discussed in an accompanying
paper.

\end{abstract}


\section{introduction}

The space-based gravitational wave detector LISA (Laser Interferometer
Space Antenna), scheduled for launch around 2011 \cite{LISA}, will
open up a window for the low frequency band below 1Hz, allowing access
to a variety of black hole sources. As one of its main targets, LISA is
expected to detect the outburst of gravitational radiation emitted during
the capture of a compact star by a supermassive black hole---a $10^{5-7}$
solar masses black hole of the kind now believed to reside in the cores
of many galaxies, including our own \cite{Nature}. Designing accurate
gravitational waveform templates for this type of astrophysical events
requires an accurate knowledge of the orbital evolution, including the
effect of radiation reaction. The evolution of such extreme mass-ratio
systems can be modeled by considering a pointlike test particle moving
in the fixed gravitational field of a black hole. One then addresses the
question of the local {\em self force} acting on this particle.
(In special cases, one may study the orbital evolution under radiation
reaction using global energy-momentum balance techniques \cite{balance}.
Such techniques, however, appear insufficient when dealing with the
astrophysically realistic case of non-equatorial eccentric orbits in Kerr
spacetime.) A general formal scheme, well physically motivated, for calculating
the gravitational self-force in curve spacetime was worked out by
Mino, Sasaki, and Tanaka (MST) \cite{MST} and, independently, by Quinn and
Wald (QW) \cite{QW}. Recently, the two groups of Barack and Ori (BO) and Mino,
Nakano, and Sasaki (MNS), have reported on a practical method for
implementing this formal scheme \cite{Letter}, allowing, for the first
time, actual calculations of the gravitational self-force for any geodesic
orbit around a Schwarzschild black hole. The purpose of the present paper
(together with the one accompanying it \cite{paperII}) is to provide a
full account of the method and results reported in \cite{Letter}.

The notion of self-forces is briefly described as follows.
Consider a pointlike particle carrying a charge $q$, which may
represent here a scalar charge, an electric charge, or a mass.
The particle is assumed to move freely in the curved background of
a black hole with mass $M\gg q$. In the limit $q\to 0$, such a particle
is known to move along a geodesic of the background geometry. However,
when endowed with a finite charge (or mass), the particle no longer
traces a background geodesic, as a result of interaction with its own
field. The finite-charge (or finite-mass) correction to the particle's
motion is then described in terms of a ``self-force'':
Treating the particle's field as a linear perturbation on the fixed black
hole background, the particle's equation of motion is written as
\begin{equation}\label{I-10}
\mu a_{\alpha}=F_{\alpha}^{\rm self},
\end{equation}
where $\mu$ is the particle's mass, $a_{\alpha}$ denotes its (covariant)
four-acceleration, and $F_{\alpha}^{\rm self}\propto O(q^2)$ describes the
leading-order self-force effect. [In the gravitational case, the
four-acceleration, as well as the (``fictitious'') self-force, are defined
through a mapping of the particle's worldline into a trajectory in the
background spacetime---see Ref.\ \cite{gauge}.]
The formal construction of $F_{\alpha}^{\rm self}$ was
described by MST \cite{MST} and QW \cite{QW} in the gravitational
case, by DeWitt and Brehme \cite{DB} and QW \cite{QW} in the electromagnetic
case, and by Quinn \cite{Q} in the scalar case. In all cases, the self-force
is constructed through
\begin{equation}\label{I-20}
F_{\alpha}^{\rm self}=\lim_{x\to z} F_{\alpha}^{\rm tail}(x)\,\,+\,\,
\text{trivial local terms},
\end{equation}
where $z$ represents a point on the particle's worldline where the
self-force is being evaluated, $x$ is a point in the neighborhood of $z$,
and the local terms are given explicitly in \cite{MST,QW,Q} (they include
the Abraham-Lorentz-Dirac force in the scalar and electromegnetic cases).
The quantity $F_{\alpha}^{\rm tail}(x)$,
the ``tail'' force, is a nonlocal contribution to the self-force,
whose occurrence reflects the essential {\em nonlocal} nature of the
radiation reaction effect in curved spacetime: waves emitted by the
particle may backscatter off spacetime curvature and later interact
back with their emitter. The tail force is formally constructed through
a worldline integral as \cite{MST,QW,Q}
\begin{equation}\label{I-30}
F_{\alpha}^{\rm tail}(x)=\lim_{\epsilon\to 0^+}
q\int_{-\infty}^{\tau_0-\epsilon}\hat\nabla_{\alpha} G[x,z(\tau)]\, d\tau,
\end{equation}
where $\tau$ is the proper time along the particle's worldline,
$\tau_0$ is the value of $\tau$ at the intersection of the worldline
with the past light cone of $x$, $G$ symbolizes a Green's function for
the particle's field, and $\hat\nabla_{\alpha}$ is a certain first-order
differential operator acting on $G$ [the explicit form of
$\hat\nabla_{\alpha}$, as well as the type of the Green's
function---whether a bi-scalar, a bi-vector or a bi-tensor---depend
on the case considered---whether scalar, electromagnetic, or
gravitational (see \cite{MST,QW,Q} for details)].
Notably, in the gravitational case, the tail force constitutes
the {\em sole} contribution to the self-force (as long as geodesics in
vacuum background are considered). It is the actual evaluation of the tail
part that has rendered practical calculations of the self-force most
challenging.

It is instructive (and later useful) to write Eq.\ (\ref{I-30}) in
the form
\begin{equation}\label{I-40}
F_{\alpha}^{\rm tail}(x)=
F_{\alpha}^{\rm full}(x)-F_{\alpha}^{\rm dir}(x),
\end{equation}
where $F_{\alpha}^{\rm full}(x)$ and $F_{\alpha}^{\rm dir}(x)$, the ``full''
and ``direct'' forces, are the quantities constructed by
replacing the integral in Eq.\ (\ref{I-30}) with
$\int_{-\infty}^{\tau_0+\epsilon}$ and
$\int_{\tau_0-\epsilon}^{\tau_0+\epsilon}$, respectively.
The ``full'' force $F_{\alpha}^{\rm full}$ is directly obtained from the
particle's ``full'' field by acting with $q\hat\nabla_{\alpha}$
[for the scalar case, e.g., see Eq.\ (\ref{III-30}) below].
The ``direct'' force $F_{\alpha}^{\rm dir}$ is the ``divergent piece''
to be removed, which is associated with the instantaneous
effect of waves propagating directly along the particle's light cone.
Note that the ``tail'' force
is hence attributed to waves scattered {\em inside} the particle's past
light-cone.

A direct implementation of the MST and QW scheme for calculating
the self-force in weak field was introduced recently by Pfenning
and Poisson \cite{Weakfield}.
To allow a practical implementation of this formal scheme
for strong-field orbits, BO devised a multipole-mode decomposition
method, relying directly on MST and QW's formal result (\ref{I-20}).
BO's {\em mode-sum method} was formulated first for the
scalar self-force \cite{MSRS-scalar}, and later for the gravitational
self-force \cite{MSRS-grav}. This method has been tested and fully
implemented for calculating the scalar self force in several cases
\cite{Burko,implementation}. The mode sum scheme (which we review in the
next section) is based on decomposing the tail force into individual
multipole-mode contributions, relating these contributions to the
``full force'' modes---which are accessible to standard
numerical analysis---and then summing over the mode contributions,
subject to a certain regularization procedure. This procedure
requires knowledge of certain analytic parameters, the ``regularization
parameters'' (RP), whose values depend on the orbit under consideration.
The RP values were derived previously for a few special orbits in
Schwarzschild spacetime: for radial and circular orbits in the scalar case
\cite{MSRS-scalar}, and for radial trajectories in the gravitational case
\cite{MSRS-grav,BL}. These (rather cumbersome) calculations were carried
out through a special local perturbative expansion of the Green's function's
multipole modes, relying directly on the integral formula (\ref{I-30}).

In this paper we present a different approach for the calculation
of the RP, based on a direct multipole decomposition of the ``direct''
piece of force. This new approach (already outlined in \cite{Letter})
allowed a rather convenient calculation of all RP values for a {\em general}
geodesic orbit in Schwarzschild spacetime, as we describe in this paper.
In particular, it provided an independent verification for the RP values
in the special cases considered previously (using the $l$-mode Green's
function analysis as mentioned above). Two variants of the new
calculation method were worked out independently by the two groups of
BO and MNS, yielding the same RP values \cite{Letter}. The calculation
by MNS has been reported in \cite{MNS}. This paper presents full
details of the RP derivation by BO.

In its basis, the calculation method presented here is applicable to
all three sorts of self-forces: scalar, electromagnetic, and gravitational.
We find it most instructive to concentrate first on the scalar case,
as a toy model. This model captures the essential parts of the calculation
technique, while avoiding several complexities and delicate issues that
show up in the gravitational and electromagnetic cases. In this paper we
thus focus on the scalar model, leaving the treatment of the gravitational
and electromagnetic cases to an accompanying paper.

It should be commented that other approaches for the calculation of
the self-force, not directly relying on the MST and QW formal scheme,
were also suggested recently. Lousto \cite{Lousto} introduced an
approach also based on a multipole decomposition but employing a
proposed zeta-function regularization scheme. Other methods were
proposed by Nakano and Sasaki \cite{NS}, Detweiler \cite{Detweiler},
and Detweiler and Whiting \cite{DW}.

The paper is arranged as follows:
In Sec.\ \ref{SecII} we review the mode-sum method,
and define the regularization parameters. The scalar toy model to
be considered in this paper is introduced in Sec.\ \ref{SecIII}.
The expression for the ``direct'' part of the scalar force is
introduced and processed in Sec.\ \ref{SecIV}, and is being formally
decomposed into modes in Sec.\ \ref{SecV}. We then prepare for
the calculation of the RP by introducing a useful coordinate system,
in Sec.\ \ref{SecVI}. The main part of our calculation is contained in
Sec.\ \ref{SecVII}, where, through an investigation of the direct force's
multipole modes, we obtain all RP values for a general trajectory in
Schwarzschild spacetime. Sec.\ \ref{SecVIII} summarizes the RP values,
and Sec.\ \ref{SecIX} provides some concluding remarks.

Throughout this paper we use geometrized units
(with $G=c=1$), and metric signature ${-}{+}{+}{+}$.

\section{reviewing the mode-sum approach}\label{SecII}

The mode-sum method was introduced in Ref.\
\cite{MSRS-scalar} for the scalar self force, and in Ref.\ \cite{MSRS-grav}
for the gravitational self-force. Here we review it using a slightly
different perspective (and notation).

In the mode-sum scheme, one first formally expands all three
quantities $F_{\alpha}^{\rm tail}(x)$, $F_{\alpha}^{\rm full}(x)$ and
$F_{\alpha}^{\rm dir}(x)$ appearing in Eq.\ (\ref{I-40}) into multipole
$l$-modes as
\begin{equation}\label{II-10}
F_{\alpha}^{\rm tail}(x)=\sum_{l=0}^{\infty}F_{\alpha}^{{\rm (tail)}l}(x),
\quad\quad
F_{\alpha}^{\rm full}(x)=\sum_{l=0}^{\infty}F_{\alpha}^{{\rm (full)}l}(x),
\quad\quad
F_{\alpha}^{\rm dir}(x)=\sum_{l=0}^{\infty}F_{\alpha}^{{\rm (dir)}l}(x)
\end{equation}
(where, recall, $x$ represents an off-worldline point in the neighborhood
of the self-force evaluation point $z$).
Here, $F_{\alpha}^{{\rm (tail)}l}$, $F_{\alpha}^{{\rm (full)}l}$, and
$F_{\alpha}^{{\rm (dir)}l}$ are the quantities obtained by summing over all
azimuthal numbers $m$ (and, in the gravitational case, also over all ten
tensor harmonics), for a given multipole number $l$.
An important benefit of the multipole decomposition is the fact that, whereas
$F_{\alpha}^{\rm full}$ and $F_{\alpha}^{\rm dir}$ both diverge at $x\to z$,
their individual modes attain finite values even at the particle's location
(though they are usually found to be discontinuous there).
Applying the multipole decomposition to Eq.\ (\ref{I-40}) we obtain
\begin{equation}\label{II-20}
F_{\alpha}^{{\rm (tail)}l}(x)=
F_{\alpha}^{{\rm (full)}l}(x)-F_{\alpha}^{{\rm (dir)}l}(x).
\end{equation}

Considering now MST and QW's expression for the self-force, Eq.\
(\ref{I-20}), we have
\begin{equation}\label{II-30}
F_{\alpha}^{\rm self}=F_{\alpha}^{\rm tail}(x=z)=
\sum_{l}F_{\alpha}^{{\rm (tail)}l}(x=z)
\end{equation}
(we hereafter ignore the trivial local terms and focus on the tail contribution).
Note that since the tail
force $F_{\alpha}^{\rm tail}(x)$ is regular at the particle's location
$z$ \cite{MST,QW}, one gets $F_{\alpha}^{\rm self}$ by just evaluating
the tail force at $x=z$.
We can then write, using Eq.\ (\ref{II-20}),
\begin{equation}\label{II-40}
F_{\alpha}^{\rm self}=\sum_{l}\lim_{x\to z}F_{\alpha}^{{\rm (tail)}l}(x)
=\sum_{l}\left[\lim_{x\to z}F_{\alpha}^{{\rm (full)}l}(x)
-\lim_{x\to z} F_{\alpha}^{{\rm (dir)}l}(x)\right],
\end{equation}
where the direction of the limit $x\to z$ is considered as {\em prescribed}.
It is important to note here that each of the two limits
$\lim_{x\to z}F_{\alpha}^{{\rm (full)}l}(x)$ and
$\lim_{x\to z}F_{\alpha}^{{\rm (dir)}l}(x)$ is, in general, directional
dependent. This, however, does not pose a problem (and the third equality
in the above chain of equalities is valid) if the direction of the limit
is prescribed: one then only has to make sure that the two limits of the
full and direct forces are taken in a consistent manner (i.e., from the
same direction).

In the last expression of Eq.\ (\ref{II-40}), the sum over $l$-modes
is guaranteed to converge [as $F_{\alpha}^{\rm tail}(x)$ is a regular
function]. However, the individual sums over the full-force modes and over
the direct-force modes usually diverge. Suppose now that one could
construct a function $h_{\alpha}^{l}$ that would make the sum
$\sum_{l}\left[\lim_{x\to z}F_{\alpha}^{{\rm (full)}l}(x)-h_{\alpha}^{l}\right]$
convergent. Then, we would have [continuing the chain of equalities
(\ref{II-40})]
\begin{eqnarray}\label{II-50}
F_{\alpha}^{\rm self}&=&
\sum_{l}\left[\left(\lim_{x\to z}F_{\alpha}^{{\rm (full)}l}(x)-h_{\alpha}^{l}
\right) -\left(\lim_{x\to z} F_{\alpha}^{{\rm (dir)}l}(x)-
h_{\alpha}^{l}\right)\right]\nonumber\\
&=&
\sum_l\left(\lim_{x\to z}F_{\alpha}^{{\rm (full)}l}(x)-h_{\alpha}^{l}\right)
-\sum_l\left(\lim_{x\to z} F_{\alpha}^{{\rm (dir)}l}(x)-h_{\alpha}^{l}\right).
\end{eqnarray}
In principle, the ``regularization function'' $h_{\alpha}^{l}$ is to be
obtained by exploring the behaviour of the full-force modes at large $l$.
However, this function can also be deduced by analyzing the large $l$
behavior of the local quantity $F_{\alpha}^{{\rm (dir)}l}$---a task
accessible to analytic treatment.
In all cases considered so far, the function $h_{\alpha}^{l}$
was found to have the general form
\begin{equation}\label{II-60}
h_{\alpha}^{l}=A_{\alpha}L+B_{\alpha}+C_{\alpha}/L,
\end{equation}
with $L\equiv l+1/2$, and where $A_{\alpha}$, $B_{\alpha}$, and $C_{\alpha}$
are $l$-independent coefficients whose values depend on the details of
the trajectory under consideration. Defining
\begin{equation}\label{II-70}
D_{\alpha}\equiv \sum_{l=0}^{\infty}
\left[\lim_{x\to z}F_{\alpha}^{{\rm (dir)}l}(x)
-A_{\alpha}L-B_{\alpha}-C_{\alpha}/L\right],
\end{equation}
we finally get from Eq.\ (\ref{II-50})
\begin{equation}\label{II-80}
F_{\alpha}^{\rm self}=\sum_{l=0}^{\infty}
\left[\lim_{x\to z}F_{\alpha}^{{\rm (full)}l}(x)
-A_{\alpha}L-B_{\alpha}-C_{\alpha}/L\right]-D_{\alpha}.
\end{equation}

Eq.\ (\ref{II-80}) constitutes the basic formula for constructing the
self-force through the mode-sum method. The four quantities
$A_{\alpha}$, $B_{\alpha}$, $C_{\alpha}$, and $D_{\alpha}$
are called the ``regularization parameters'' (RP). The full modes
$F_{\alpha l}^{\rm full}$, recall, are directly obtained from the
``full'' field modes [see Eq.\ (\ref{III-30}) below for the construction
of the full force in the scalar case], which, in turn, are calculated
using standard numerical techniques. Eq.\ (\ref{II-80}) thus describes a
practical scheme for constructing the self force, given the values of the RP.

In this paper (dealing with the scalar self-force) and in the accompanying
paper (dealing with the gravitational and electromagnetic self-forces)
we derive the values of all RP needed for implementing Eq.\ (\ref{II-80})
for any geodesic orbit in Schwarzschild spacetime.

\section{Scalar toy-model}\label{SecIII}

We consider a particle of a scalar charge $q$, moving freely
in the vacuum exterior of a Schwarzschild black hole with mass $M\gg q$.
In the lack of self-force, the particle moves along a geodesic
$z^{\mu}(\tau)$ with specific energy and angular momentum parameters
$\cal E$ and $\cal L$, respectively. We shall consider the self-force
acting on the particle at a point along its worldline which we
denote by $z\equiv (t_{0},r_{0},\theta_{0},\varphi _{0})$ (where
$t,r,\theta,\varphi$ are the standard Schwarzschild coordinates).
Let also $x\equiv (t,r,\theta,\varphi)$ denote a point in the close
neighborhood of $z$.

The particle induces a scalar field $\Phi^{\rm full}(x)$, which we
shall treat as a linear perturbation over the fixed Schwarzschild
background. In our model, the field $\Phi^{\rm full}(x)$ is assumed
to satisfy the (minimally-coupled) Klein--Gordon equation
\begin{equation}\label{III-10}
\Box\Phi^{\rm full}\equiv\Phi^{{\rm full};\alpha}_{;\alpha}=-4\pi\rho,
\end{equation}
where a semicolon denotes covariant differentiation with respect to the
background geometry, and the scalar charge density is given by
\begin{equation}\label{III-20}
\rho(x)=q\int_{-\infty}^{\infty}\delta^4[x-z(\tau)](-g)^{-1/2}d\tau
\end{equation}
($g$ being the metric determinant).
We now define the ``full force'' as the vector field
\begin{equation}\label{III-30}
F^{\rm full}_{\alpha}(x)\equiv q\Phi^{\rm full}_{,\alpha}.
\end{equation}
Note that both the full field $\Phi^{\rm full}(x)$ and the full force
$F^{\rm full}_{\alpha}(x)$ obviously diverge on the worldline, but are
otherwise well defined.

The force definition (\ref{III-30}) complies with Quinn's definition
\cite{Q}. It differs from the expression used by MNS, which involves
a spatial projection of the scalar force [see Eq.\ (1.3) in \cite{MNS}].
We prefer to adopt here the force definition (\ref{III-30}) for
several reasons:
(i) It is a simpler definition, which nevertheless serves
as an effective toy-model for the realistic gravitational case.
(ii) It avoids the need to consider an off-worldline extension of
the four-velocity, as necessary for defining the spatially-projected
force.
(iii) The force model (\ref{III-30}) is naturally derived from a
Lagrangian formalism, and is hence consistent with global stress-energy
conservation---unlike the spatially projected force \cite{Q,QW99}.

Finally, we introduce the notions of the ``direct'' field $\Phi^{\rm dir}$
and the ``tail'' field $\Phi^{\rm tail}=\Phi^{\rm full}-\Phi^{\rm dir}$
(see \cite{Q,MNS}), from which the direct and tail forces are derived by
\begin{equation}\label{III-40}
F^{\rm dir}_{\alpha}(x)= q\Phi^{\rm dir}_{,\alpha},
\quad\quad
F^{\rm tail}_{\alpha}(x)= q\Phi^{\rm tail}_{,\alpha}.
\end{equation}
Recall that the ``direct'' field is the part of the scalar field
propagated directly along the particle's light cone, while the ``tail''
part is associated with reflections of the field {\em inside} the light
cone.

\section{Direct force: preliminaries} \label{SecIV}

The form of the direct scalar field $\Phi^{\rm dir}$ was worked out
by MNS \cite{MNS} (see also some preliminary results in \cite{MN}),
by studying the Hadamard expansion of the field equation.
Let $\epsilon(x)$ denote the spatial geodesic distance from the point
$x$ to the geodesic $z(\tau)$ (i.e., the length of the short geodesic
section connecting $x$ to the worldline and normal to it),
and let $\delta x^{\mu}\equiv x^{\mu}-z^{\mu}$. Then, the direct
scalar field obtained by MNS can be written in the form
\begin{equation}\label{IV-10}
\Phi^{\rm dir}(x)=\frac{q\,\hat f(\delta x)}{\epsilon}+{\rm const},
\end{equation}
where $\hat f$ is a regular function of $\delta x$ (and $z$)
satisfying
\begin{equation}\label{IV-15}
\hat f=1+O(\delta x^2)
\end{equation}
(the explicit form of $f$ will not be needed in the analysis below).
Introducing the squared geodesic distance $S(\delta x)\equiv \epsilon^2$,
the direct scalar force is then given by
\begin{equation}\label{IV-20}
F_{\alpha}^{\rm dir}(x)= q\Phi_{,\alpha}^{\rm dir}
=q^2\left[\hat f_{,\alpha}S^{-1/2}-(\hat f/2)S^{-3/2}S_{,\alpha}\right].
\end{equation}

Consider now the Taylor expansion of the function $S(\delta x)$ about
$\delta x=0$. We write this expansion as
\begin{equation}\label{IV-25}
S=S_0+S_1+S_2+\cdots
\end{equation}
where $S_0,S_1,\ldots$ represent terms of homogeneous orders
$\delta x^2,\delta x^3,\ldots$, respectively. Note that this decomposition
of $S$ is no longer covariant, and the individual terms $S_n$ will depend
on the choice of coordinate system.
Below we shall need only the two leading terms, $S_0$ and $S_1$, which we
obtain in Appendix \ref{AppA}. We find
\begin{mathletters} \label{IV-30}
\begin{equation} \label{IV-30a}
S_0=(g^0_{\mu\nu}+u_{\mu}u_{\nu})\delta x^{\mu}\delta x^{\nu},
\end{equation}
\begin{equation} \label{IV-30b}
S_1=\left(u_{\lambda}u_{\gamma}\Gamma^{\lambda0}_{\alpha\beta}+
g^0_{\alpha\beta,\gamma}/2 \right)
\delta x^{\alpha}\delta x^{\beta}\delta x^{\gamma},
\end{equation}
\end{mathletters}
where $u^{\alpha}\equiv dx^{\alpha}/d\tau$ is the four-velocity at $z$,
and $\Gamma^{\lambda 0}_{\alpha\beta}$ and $g^0_{\alpha\beta}$
denote, respectively, the connection coefficients and metric functions
evaluated at $\delta x=0$ (namely, at $x=z$).
Substituting Eqs.\ (\ref{IV-15}) and (\ref{IV-30}) in Eq.\ (\ref{IV-20}),
we now obtain a Taylor expansion for the direct force, which we may
expresse as
\begin{equation}\label{IV-40}
F_{\alpha}^{\rm dir}(x)=q^2\left[
\epsilon_0^{-3} P_{\alpha}^{(1)} +
\epsilon_0^{-5}P_{\alpha} ^{(4)} +
\epsilon_0^{-7}P_{\alpha}^{(7)} + \cdots\right].
\end{equation}
Here, $\epsilon_0\equiv S_0^{1/2}$, and $P_{\alpha}^{(n)}$ denote terms of
homogeneous order $O(\delta x^n)$. Note that the term of the form
$\propto\epsilon^{-1}\delta x$ in Eq.\ (\ref{IV-20}) can be written
as $\propto\epsilon^{-7}(\epsilon^6\delta x)\propto\epsilon^{-7}\delta x^7$
and then be absorbed in the term $\epsilon_0^{-7}P_{\alpha}^{(7)}$.
Similarly, terms of the form $\propto\epsilon^{-3}\delta x^2$ may be
expressed as $\propto\epsilon^{-5}\delta x^{4}$ and be absorbed in
$\epsilon_0^{-5}P_{\alpha}^{(4)}$, and so on.
Note also that the three terms presented in Eq.\ (\ref{IV-40}) are of
orders $\delta x^{-2}$, $\delta x^{-1}$, and $\delta x^{0}$, respectively.
The three dots ($\cdots$) in that equation represent terms that vanish
in the limit $\delta x\to 0$ [like, e.g., $\epsilon_0^{-9} P_{\alpha}^{10}
\propto O(\delta x)$]. In the following analysis we shall need
the explicit values of only $P_{\alpha}^{(1)}$ and $P_{\alpha}^{(4)}$,
which are given by
\begin{mathletters}\label{IV-50}
\begin{equation}\label{IV-50a}
P_{\alpha}^{(1)}=-\frac{1}{2}S_{0,\alpha},
\end{equation}
\begin{equation}\label{IV-50b}
P_{\alpha}^{(4)}=-\frac{1}{2}S_{0}S_{1,\alpha}+\frac{3}{4}S_{0,\alpha }S_1.
\end{equation}
\end{mathletters}

Now, in constructing the self-force, one is merely concerned with the
behavior of the direct force at $x\to z$---see, e.g, Eq.\ (\ref{II-40}).
Thus, the terms represented by the three dots ($\cdots$) in Eq.\ (\ref{IV-40}),
which vanish in the limit $x\to z$, are irrelevant for calculating the
self-force, and may be ignored in our analysis. We hence introduce a
``revised'' version of the direct-force by omitting these terms (retaining,
though, the notation $F_{\alpha}^{\rm dir}$):
\begin{equation}\label{IV-60}
F^{\rm dir}_{\alpha}=q^2\left[F_{\alpha}^{(A)}+F_{\alpha}^{(B)}
+F_{\alpha}^{(C)}\right],
\end{equation}
where
\begin{equation}\label{IV-70}
F_{\alpha}^{(A)}\equiv \epsilon_0^{-3}P_{\alpha}^{(1)},\quad\quad
F_{\alpha}^{(B)}\equiv \epsilon_0^{-5}P_{\alpha}^{(4)},\quad\quad
F_{\alpha}^{(C)}\equiv \epsilon_0^{-7}P_{\alpha}^{(7)}.
\end{equation}
Note that this splitting of $F^{\rm dir}_{\alpha}$ holds for any choice of
coordinates $x^{\mu}$ which are sufficiently regular in the neighborhood of
$z$ (though the coefficients $P_{\alpha}^{(n)}$ will depend on the choice of
coordinates).

\section{Multipole decomposition} \label{SecV}

Next, we consider the multipole decomposition of $F_{\alpha}^{\rm dir}$.
Let
\begin{equation}\label{V-10}
F_{\alpha}^{\rm dir}(x)=
\sum_{lm}F_{\alpha}^{lm}(r,t)Y^{lm}(\theta,\varphi),
\end{equation}
where
$Y^{lm}(\theta,\varphi)$ are spherical harmonics.
We denote by $F^l_{\pm\alpha}$ the total $l$-mode contribution to
the direct force at $z$:
\begin{equation}\label{V-20}
F_{\pm\alpha}^l\equiv \lim_{\delta r\to 0^{\pm}}
\sum_m F_{\alpha}^{lm}(r,t_0)Y^{lm}(\theta_0,\varphi_0).
\end{equation}
Note that $F_{\pm\alpha}^l=\lim_{x\to z}F_{\alpha}^{({\rm dir})l}(x)$ [as in
Eq.\ (\ref{II-40}), e.g.], where the direction of the limit is explicitly
specified such that $x$ approaches $z$ ``from the radial direction''.
The $\pm$ sign corresponds to the two possible radial limits,
$r\to r_0^+$ or rather $r\to r_0^-$. This choice, of taking the radial
limit, appears most convenient in our multipole-mode scheme. In
particular, it is most easily implemented in the (numerical) calculation
of the full force modes (recall that the limit $x\to z$ of both the
direct and full forces must be taken from the same direction).

Eq.\ (\ref{V-20}) is invariant under rotation in the subspace of angular
coordinates $\theta,\varphi$. We take advantage of this property,
and re-define the angular coordinates such that $z$ is located
at the pole, i.e., $\theta_0=0$. Due to angular-momentum conservation, the
particle is now confined to move on a plane of constant $\varphi$, which we
take as $\varphi=0,\pi$ (the value of the $\varphi$ coordinate is fixed along
the particle's trajectory, apart from a ``jump'' at the two poles
$\theta=0,\pi$). The particle's four-velocity now satisfies
$u^{\varphi}=0$.

The above setup is beneficial in that the $l$-mode $F_{\pm\alpha}^l$
is now composed of only the axially-symmetric $m=0$ harmonic: Recall
that $Y^{lm}$ vanishes at $\theta=0$ for any $m\ne 0$, and
$Y^{l,m=0}(\theta=0)=[L/(2\pi)]^{1/2}P_l(1)$, where $P_l(\cos\theta)$
is the Legendre polynomial and $P(1)=1$.
Consequently, we find from Eq.\ (\ref{V-20})
\begin{equation}\label{V-30}
F_{\pm\alpha}^l=\lim_{\delta r\to 0^{\pm}} [L/(2\pi)]^{1/2}
F_{\alpha}^{l,m=0}(r,t_0).
\end{equation}
The mode $F_{\alpha}^{l,m=0}$ is given by the integral
\begin{eqnarray}\label{V-40}
F_{\alpha}^{l,m=0}(r,t)&=&
\int F_{\alpha}^{\rm dir}(r,t,\theta,\varphi)[Y^{l,m=0}]^*d\Omega
\nonumber\\
&=&[L/(2\pi)]^{1/2}
\int F_{\alpha}^{\rm dir}(r,t,\theta,\varphi)P_l(\cos\theta)\,d\Omega,
\end{eqnarray}
where $d\Omega\equiv d\cos\theta\, d\varphi$ and the asterisk denotes
complex conjugation.
Combining Eqs.\ (\ref{V-30}) and (\ref{V-40}) we finally obtain
the following integral expression for the total $l$-mode direct force:
\begin{equation}\label{V-50}
F_{\pm\alpha}^l =
\lim_{\delta r\to 0^{\pm}}
\frac{L}{2\pi}\int F_{\alpha}^{\rm dir}(r,t_0,\theta,\varphi)
P_l(\cos\theta)\,d\Omega.
\end{equation}

\section{Regular coordinate system}\label{SecVI}

The coordinate system ($t,r,\theta,\varphi$) is singular at
$\theta=\theta_0=0$. This singularity makes the expansion (\ref{IV-25}),
(\ref{IV-30}) inapplicable in these coordinates. To overcome this difficulty,
we introduce the two ``locally-cartesian angular coordinates''
\begin{equation}\label{VI-10}
x=\rho(\theta)\cos\varphi,\quad  y=\rho(\theta)\sin\varphi,
\end{equation}
where $\rho(\theta)$ is a sufficiently regular, odd function of
$\theta $, admitting the expansion
\begin{equation}\label{VI-20}
\rho (\theta)=\theta+\rho_1\theta^3+\rho_2\theta^5+\cdots.
\end{equation}
For later convenience we shall also demand that $\rho(\theta)$ grows
monotonously within the entire domain $0\leq \theta<\pi$, such that
$\rho(\theta)$ is invertible. [An obvious natural choice would be
$\rho=\theta$; however, later we shall make the specific choice
$\rho(\theta)=2\sin(\theta/2)$ which will simplify our calculations.]

Using the relations
$d\rho/d\theta=1+\rho^2 h_{1}(\rho^2)$ and
$\rho^2/\sin^2\theta=1+\rho^2 h_2(\rho^2)$
[easily followed from the above definition of $\rho(\theta)$],
where $h_1$ and $h_2$ are both regular functions of $\rho^2$, one finds
that the contravariant components of the metric tensor now
take the form
\begin{eqnarray}\label{VI-30}
g^{xx}&=&r^{-2}(1+x^{2}h_{1}+y^{2}h_{2}), \nonumber\\
g^{yy}&=&r^{-2}(1+y^{2}h_{1}+x^{2}h_{2}), \nonumber\\
g^{xy}&=&r^{-2}(h_{1}-h_{2})xy.
\end{eqnarray}
The point $z$ is located at $x=y=0$. The above tensor $g^{\alpha \beta }$
is perfectly regular in the neighborhood of this point---and so is the
covariant metric $g_{\alpha \beta }$.
In the particle's location itself, $x=y=0$, the line element takes the
simple form
\begin{equation}\label{VI-40}
g_{xx}^{0}=g_{yy}^{0}=r_0^2,\quad\quad g_{xy}^{0}=0
\end{equation}
[along with
$g_{tt}^{0}=-(1-2M/r_0)$ and $g_{rr}^{0}=(1-2M/r_0)^{-1}$].

Note that the particle's geodesic is confined to $y=0$ and correspondingly
$u^{y}=0$. Also, since $z$ is located at $x=y=0$, we have
$\delta x^{x}=x,$ $\delta x^{y}=y$. Finally, we comment that the particle's
angular momentum is given as ${\cal L}=u_x$ {\em evaluated at $z$}
(but note that $u_x$ is {\em not} conserved along the geodesic).

\section{Investigating the $\lowercase{l}$-mode of the direct force}
\label{SecVII}

\subsection{Is the $x\to z$ limit interchangeable with the
Legendre integral?}

We now explore in more detail the $l$-mode direct force $F^l_{\pm\alpha}$,
based on the integral formula (\ref{V-50}). Recalling that the direct
force itself is composed of three terms, Eq.\ (\ref{IV-60}), we write
\begin{equation}\label{VII-10}
F^l_{\pm\alpha}=q^2\left[
F_{\alpha}^{(A)l}+F_{\alpha}^{(B)l}+F_{\alpha}^{(C)l}\right],
\end{equation}
where $F_{\alpha}^{(W)l}$ ($W$ standing for $A$, $B$, or $C$) denotes the
contribution to $F^l_{\pm\alpha}$ [through Eq.\ (\ref{V-50})] due to the
term $F_{\alpha}^{(W)}$ of the direct force :
\begin{equation}\label{VII-20}
F_{\pm\alpha}^{(W)l} =\lim_{\delta r\to 0^{\pm}}
\frac{L}{2\pi}\int F_{\alpha}^{(W)}(r,t_0,\theta,\varphi)
P_l(\cos\theta)\,d\Omega.
\end{equation}
Recall that the various terms $F_{\alpha}^{(W)}$ are given in Eq.\
(\ref{IV-70}).

The task of evaluating the various contributions $F_{\pm\alpha}^{(W)l}$
would be much simplified could we interchange the limit $\delta r\to 0^{\pm}$
and the integration in Eq.\ (\ref{VII-20}). Is such an interchange allowed?
In Appendix \ref{AppB} we address this question, and show that interchanging
the $\delta r\to 0^{\pm}$ limit and the Legendre integral is indeed allowed
for $W=B$ and $W=C$; however, as evident from the explicit calculation below,
such an interchange is no valid for $W=A$. Here we present a heuristic
argument suggesting why the interchange is valid for $W=B,C$, and why
it might fail for $W=A$. A sketch of a mathematical proof is provided
in Appendix \ref{AppB}.

For a given small separation $\delta x=x-z$, assume that all components
of $\delta x^{\alpha}$ scale as $\delta r$ (we assume $\delta r\ne 0$).
Since $\epsilon_0$ then scales like $\delta r$ too, we find that the
various terms $F_{\alpha}^{(W)}$ scale as
\begin{equation}\label{VII-30}
F_{\alpha}^{(A)}=\epsilon_0^{-3}P_{\alpha}^{(1)}\propto \delta r^{-2},
\quad\quad
F_{\alpha}^{(B)}=\epsilon_0^{-5}P_{\alpha}^{(4)}\propto \delta r^{-1},
\quad\quad
F_{\alpha}^{(C)}=\epsilon_0^{-7}P_{\alpha}^{(7)}\propto \delta r^{0},
\end{equation}
where the proportion coefficients only depend on the ``direction'' of
$\delta x^{\alpha}$ (i.e., on the ratios between its various components).
To consider the interchangeability of the limit and integral in Eq.\
(\ref{VII-20}), one is mainly concerned with the contribution to the
integral from small $x,y$ values (i.e., from the immediate neighborhood
of the integrand's singular point $z$). To find out how this small piece
of integral scales with $\delta r$, we consider the small integration
area around $z$, in the $xy$ plain, defined by
$\rho=(x^2+y^2)^{1/2}<\delta r$ (for a given $\delta r\ne 0$).
Observing that this integration area scales like $\delta r^{2}$
and relying on the scale relations (\ref{VII-30}), one finds
that this small-$\delta r$ contribution to the integral scales
like $\delta r^0$ for $F_{\pm\alpha}^{l(A)}$, like $\delta r^1$
for $F_{\pm\alpha}^{l(B)}$, and like $\delta r^2$ for $F_{\pm\alpha}^{l(C)}$.
Namely, upon taking the limit $\delta r\to 0$, the small-$\delta r$ piece of
integration vanishes for $W=B,C$, but not for $W=A$. This suggests that
we may interchange the limit and integration for $W=B,C$, but not for $W=A$.
A more rigorous mathematical treatment implies that this is indeed the
case (see Appendix \ref{AppB}).

We are thus allowed to write
\begin{equation}\label{VII-40b}
F_{\alpha}^{l(B,C)}=
\frac{L}{2\pi}\int F_{\alpha}^{(B,C)}(r_0,t_0,x,y)P_l(\cos\theta)d\Omega.
\end{equation}
However, for $W=A$ we must use the original expression:
\begin{equation}\label{VII-40a}
F_{\pm\alpha}^{l(A)}=
\lim_{\delta r\to 0^{\pm}}
\frac{L}{2\pi}\int F_{\alpha}^{(A)}(r,t_0,x,y)P_l(\cos\theta)d\Omega.
\end{equation}
For later convenience we give here explicitly the form of
$F_{\alpha }^{(B,C)}$ for $r=r_0,t=t_0$. We have
\begin{equation}\label{VII-50}
F_{\alpha}^{(B)}=\hat\epsilon_0^{-5}P_{\alpha}^{(4)}(x,y),
\quad\quad
F_{\alpha}^{(C)}=\hat\epsilon_0^{-7}P_{\alpha}^{(7)}(x,y),
\end{equation}
where $P_{\alpha}^{(n)}(x,y)$ is a polynomial of homogeneous
order $n$ in $x$ and $y$, and $\hat\epsilon_0$ is the reduction of
$\epsilon_0$ to $\delta r=\delta t=0$: We find, recalling
$\rho^2=x^2+y^2$ and $u_y=0$,
\begin{equation}\label{VII-60}
\hat{\epsilon}_{0}=\left(r_0^2\rho^2+u_x^2 x^2\right)^{1/2}.
\end{equation}
Note that $\hat\epsilon_0$ is an even function of both $x$ and
$y$---a fact that will play a crucial role in the analysis below.

\subsection{Calculating $F_{\alpha}^{l(C)}$}

Let us first evaluate the term $F_{\alpha}^{l(C)}$.
We observe that the integrand in Eq.\ (\ref{VII-40b}) is composed
of three factors:
$\hat\epsilon_0^{-7}\times P_{\alpha}^{(7)}(x,y)\times P_l(\cos\theta)$.
Since $\hat{\epsilon}_{0}$ and $\cos\theta(\rho)$ are even function of
both $x$ and $y$, then so are the factors
$\hat\epsilon_0^{-7}$ and $P_l(\cos\theta)$.
However, each of the eight terms of $P_{\alpha}^{(7)}(x,y)$ (proportional
to $x^7y^0,x^6y,\ldots, x^0y^7$) is of odd power in either $x$ or $y$.
Hence, the overall integrand in Eq.\ (\ref{VII-40b}) is composed only
of terms which are odd in either $x$ or $y$. As a consequence, the
integral is found to vanish identically, yielding
\begin{equation}\label{VII-70}
F_{\alpha}^{l(C)}=0.
\end{equation}

\subsection{Calculating $F_{\alpha}^{l(B)}$}

We next turn to consider $F_{\alpha}^{l(B)}$.
The integrand in Eq.\ (\ref{VII-40b}) now takes the form
$\hat\epsilon_0^{-5}\times P_{\alpha}^{(4)}(x,y)\times P_l(\cos\theta)$.
Note that the polynomial $P_{\alpha}^{(4)}(x,y)$ may now
contain terms which are even in both $x$ and $y$, yielding,
in general, a non-vanishing contribution to the integral.
To proceed, one thus has to be provided with the explicit form
of $P_{\alpha}^{(4)}$.

The explicit form of the polynomial $P_{\alpha}^{(4)}(x,y)$ is
obtained by substituting for $S_0$ and $S_1$ (and their gradients)
from Eqs.\ (\ref{IV-30}) in Eq.\ (\ref{IV-50b}), taking
$\delta r=\delta t=0$ and recalling $\delta x=x$, $\delta y=y$,
and $u_y=0$. One thereby obtains
\begin{mathletters}\label{VII-80}
\begin{equation}\label{VII-80a}
P_{\alpha}^{(4)}(x,y)=P^{(x)}_{\alpha}\,x^4+P^{(xy)}_{\alpha}\,x^2y^2
+P^{(y)}_{\alpha}\,y^4,
\quad\quad {\rm for\ } \alpha=t,r,x,
\end{equation}\
\begin{equation}\label{VII-80b}
P_y^{(4)}(x,y)=P^{(x)}_{y}\,x^3y +P^{(y)}_{y}\,xy^3,
\end{equation}
\end{mathletters}
where the various coefficients are explicitly given by
\begin{eqnarray}\label{VII-90}
P^{(x)}_{r}&=&-\frac{1}{2}\left[f^{-1}\dot{r}^2r_0(2u_x^2-r_0^2)
+r_0^{-1}(2u_x^4+3u_x^2r_0^2+r_0^4)\right],
\nonumber\\
P^{(xy)}_{r}&=&-\frac{1}{2}r_0\left[3u_x^2+2r_0^2+2f^{-1}\dot{r}^2
(u_x^2-r_0^2)\right],
\nonumber\\
P^{(y)}_{r}&=&-\frac{1}{2}r_0^3(1-f^{-1}\dot{r}^2),
\end{eqnarray}
\begin{equation}\label{VII-100}
P^{(x)}_{t}=-r_0 u_t\dot{r}(u_x^2-r_0^2/2),
\quad\quad
P^{(xy)}_{t}=-r_0 u_t\dot{r}(u_x^2-r_0^2),
\quad\quad
P^{(y)}_{t}=\frac{1}{2}r_0^3 u_t\dot{r},
\end{equation}
\begin{equation}\label{VII-110}
P^{(x)}_{x}=0,
\quad\quad
P^{(xy)}_{x}=\frac{1}{2}r_0u_x\dot{r}(r_0^2-2u_x^2),
\quad\quad
P^{(y)}_{x}=\frac{1}{2}r_0^3u_x\dot{r},
\end{equation}
\begin{equation}\label{VII-120}
P^{(x)}_{y}=r_0 u_x\dot{r}(u_x^2-r_0^2/2),
\quad\quad
P^{(y)}_{y}=-\frac{1}{2}r_0^3 u_x\dot{r}.
\end{equation}
In these expressions $f\equiv (1-2M/r_0)$, $\dot r\equiv u^r$,
and all four-velocity components are evaluated at $z$.
Note that the components $P_{t}^{(4)}$, $P_{r}^{(4)}$, and $P_{x}^{(4)}$
consist of only terms which are {\em even} in both $x$ and $y$. On the
other hand, the $y$ component $P_{y}^{(4)}$ contains only terms which
are {\em odd} in both coordinates.

Consider first the $y$ component: Both terms $\propto x^3y$ and
$\propto xy^3$ of the polynomial $P_y^{(4)}$ yield, upon integrating,
no contribution to $F_{y}^{l(B)}$, and one immediately obtains
\begin{equation}\label{VII-130}
F_{y}^{l(B)}=0.
\end{equation}
The other components of $F_{\alpha}^{l(B)}$ do not similarly vanish:
Recalling $x=\rho\cos\varphi$ and $y=\rho\sin\varphi$, and
expressing $\hat\epsilon_0$ in the form $\hat\epsilon_0=r_0\rho(\theta)
(1+r_0^{-2}u_x^2\cos^2\varphi)^{1/2}$, we may write the double
integral in Eq.\ (\ref{VII-40b}) in the factorized form
\begin{equation}\label{VII-140}
F_{\alpha}^{l(B)}=r_{0}^{-5}I^{\theta}\,I^{\varphi}_{\alpha},
\end{equation}
where
\begin{mathletters}\label{VII-150}
\begin{equation}\label{VII-150a}
I^{\theta}\equiv \frac{L}{2\pi}\int_{-1}^{1}
\frac{P_l(\cos\theta)}{\rho(\theta)}\,d\cos\theta,
\end{equation}
\begin{equation}\label{VII-150b}
I_{\alpha}^{\varphi}\equiv \int_0^{2\pi}
\frac{P_{\alpha}^{(x)}\cos^4\varphi+ P_{\alpha}^{(xy)}\cos^2
\varphi\sin^2\!\varphi+ P_{\alpha}^{(y)}\sin^4\varphi}
{(1+r_0^{-2}u_{x}^{2}\cos^2\varphi)^{5/2}}\,d\varphi.
\end{equation}
\end{mathletters}
We now take advantage of the freedom we still have in specifying
the function $\rho(\theta)$, and make the convenient choice
\begin{equation}\label{VII-160}
\rho=2\sin(\theta/2).
\end{equation}
With this choice, the integral $I^{\theta}$ becomes a standard one,
reading simply
\begin{equation}\label{VII-170}
I_{\theta}=(2\pi)^{-1}
\end{equation}
[see, e.g., Eq.\ (7.225-3) of \cite{GR}].
The integral $I_{\alpha}^{\varphi}$, in turn, is a linear combination
of standard elliptic integrals. It can be expressed as
\begin{equation}\label{VII-180}
I_{\alpha}^{\varphi}=a_{\alpha}\hat K(w)+b_{\alpha}\hat E(w),
\end{equation}
where $\hat{K}(w)$ and $\hat{E}(w)$ are two complete elliptic integrals
of the first and second kinds, respectively, the argument $w$ is given by
\begin{equation}\label{VII-190}
w\equiv \frac{u_x^2}{r_0^2+u_x^2},
\end{equation}
and the coefficients $a_{\alpha}$ and $b_{\alpha}$ read
\begin{eqnarray}\label{VII-200}
a_{\alpha}&=&-\frac{4}{3}(r_0/u_x)^3w^{-1/2}
\left[a^{(x)} P_{\alpha}^{(x)}+a^{(xy)}P_{\alpha}^{(xy)}
+a^{(y)} P_{\alpha}^{(y)}\right],
\nonumber\\
b_{\alpha}&=&-\frac{4}{3}(r_0/u_x)w^{-3/2}
\left[b^{(x)} P_{\alpha}^{(x)}+b^{(xy)} P_{\alpha}^{(xy)}
+b^{(y)} P_{\alpha}^{(y)}\right],
\end{eqnarray}
with
\begin{eqnarray}\label{VII-210}
a^{(x)} &=& (w+2)(w-1),    \nonumber\\
a^{(xy)}&=& -2(w-1),       \nonumber\\
a^{(y)} &=& (3w-2),        \nonumber\\
b^{(x)} &=& 2(w-1)^2(w+1), \nonumber\\
b^{(xy)}&=& -(w-2)(w-1),    \nonumber\\
b^{(y)} &=& 2(1-2w).
\end{eqnarray}

The explicit form of the desired contribution $F_{\alpha}^{l(B)}$
(for $\alpha=r,t,x$) is finally obtained by inserting the values of
$P_{\alpha}^{(x)}$, $P_{\alpha}^{(xy)}$, and $P_{\alpha}^{(y)}$
[given in Eqs.\ (\ref{VII-90})--(\ref{VII-120})] in the above expressions
for $a_{\alpha}$ and $b_{\alpha}$, constructing $I_{\alpha}^{\varphi}$
through Eq.\ (\ref{VII-180}), and substituting in Eq.\ (\ref{VII-140}).
This yields
\begin{mathletters}\label{VII-220}
\begin{equation}\label{VII-220a}
F_{r}^{l(B)}=\frac{1}{r_0^2}\,
\frac{(\dot{r}^2-2u_t^2)\hat{K}(w)+(\dot{r}^2+u_t^2)\hat{E}(w)}
{\pi f V^{3/2}},
\end{equation}
\begin{equation}\label{VII-220b}
F_{t}^{l(B)}=\frac{1}{r_0^2}\,\frac{u_t\dot{r}[\hat{K}(w)-2\hat{E}(w)]}
{\pi V^{3/2}},
\end{equation}
\begin{equation}\label{VII-220c}
F_{x}^{l(B)}=\frac{1}{r_0}\,\frac{\dot{r}[\hat{K}(w)-\hat{E}(w)]}
{\pi (u_x/r_0)V^{1/2}},
\end{equation}
\end{mathletters}
where $V\equiv 1+u_x^2/r_0^2$.

Note the remarkable fact that the contribution $F_{\alpha}^{l(B)}$
is {\em independent of $l$}.

\subsection{Calculating $F_{\pm\alpha}^{l(A)}$}

Finally, let us evaluate $F_{\pm\alpha}^{l(A)}$.
Recalling $F_{\alpha}^{(A)}=\epsilon_0^{-3}P_{\alpha}^{(1)}$
and using Eqs.\ (\ref{IV-50a}) and (\ref{IV-30a}),
Eq.\ (\ref{VII-40a}) becomes
\begin{equation}\label{VII-230}
F_{\pm\alpha}^{l(A)}= -[L/(2\pi)](g^0_{\alpha\beta}+u_{\alpha}u_{\beta})
\tilde F_{\pm}^{l\beta},
\end{equation}
where
\begin{equation}\label{VII-240}
\tilde F_{\pm}^{l\beta}\equiv \lim_{\delta r\to 0^{\pm}}
\int \delta x^{\beta}\epsilon_0^{-3} P_l(\cos\theta)d\Omega.
\end{equation}
Note that we have already taken here the limit $\delta t\to 0$, hence
the integrand ($\propto \delta x^{\beta}$) vanishes identically for
$\beta=t$. Also, since $\cos\theta(\rho)$ and $\epsilon_0$, given
explicitly by
\begin{equation}\label{VII-250}
\epsilon_0=\left[r_0^2\rho^2
+g_{rr}^0\delta r^{2}+(u_{r}\delta r+u_{x}x)^2\right]^{1/2},
\end{equation}
are both even functions of $y$, the integral in Eq.\ (\ref{VII-240})
obviously vanishes for $\beta=y$. Hence,
\begin{equation}\label{VII-260}
\tilde F_{\pm}^{lt}=\tilde F_{\pm}^{ly}=0.
\end{equation}

Consider now Eq.\ (\ref{VII-240}) for the two remaining components,
$\beta=r,x$. First, we change the integration variables to $x,y$.
Since the Jacobian is $\partial(\theta,\varphi )/\partial(x,y)=
(\rho\rho')^{-1}$ (where $\rho'=d\rho/d\theta$), Eq.\ (\ref{VII-240}) becomes
\begin{equation}\label{VII-265}
\tilde F_{\pm}^{l\beta}\equiv \lim_{\delta r\to 0^{\pm}}
\int \delta x^{\beta}\epsilon_0^{-3}H(\rho)dxdy,
\end{equation}
where $H(\rho)\equiv P_l(\cos\theta)\sin\theta(\rho\rho')^{-1}$.
The function $H(\rho)$ is a regular, even function of $\theta$
(and of $\rho$), with $H(0)=1$. We thus write it as
$H(\rho)=1+\rho^2 \hat H(\rho)$, where the function $\hat H(\rho)$
admits a regular (even) Taylor expansion at $\rho=0$. Accordingly,
we divide $\tilde F_{\pm}^{l\beta}$ into two contributions,
\begin{equation}\label{VII-270}
\tilde F_{\pm}^{l\beta}=\lim_{\delta r\to 0^{\pm}}
(I^{\beta}_1+I^{\beta}_2),
\end{equation}
where
\begin{eqnarray}\label{VII-280}
I^{\beta}_1 &\equiv & \int \delta x^{\beta}\epsilon_0^{-3} dxdy,
\nonumber\\
I^{\beta}_2 &\equiv &
\int \delta x^{\beta}\epsilon_0^{-3}\rho^2 \hat H(\rho) dxdy.
\end{eqnarray}

Consider first the contribution $I^{\beta}_2$:
Near $\rho=0$, the integrand in this term scales like $\delta r^0$,
thus the integrated singular contribution scales like $\delta r^2$.
Hence, based on precisely the same argument applied in Appendix
\ref{AppB} with regard to the term $F_{\alpha}^{(C)}$,
we find that the integral $I^{\beta}_2$ is sufficiently regular to
allow us interchanging the orders of the $\delta r\to 0$ limit and
integration:
\begin{equation}\label{VII-290}
\lim_{\delta r\to 0^{\pm}} I^{\beta}_2=I^{\beta}_2(\delta r=0).
\end{equation}
Doing so, we find that the contribution from $I^{\beta}_2$ to
$\tilde F_{\pm}^{l\beta}$ vanishes for either $\beta=r$ or $\beta=x$:
For $\beta=r$, the integrand vanishes identically; for $\beta=x$, the
integrand, evaluated at $\delta r=0$, becomes an odd function of $x$
[see Eq.\ (\ref{VII-60})], which vanishes upon integrating.

To calculate the remaining contribution $I^{\beta}_1$, we divide the
domain of integration in Eq.\ (\ref{VII-280}) into two regions:
Let $H^{\rm in}$ denote the square $-h<x,y<h$, for some particular
$0<h<1$ (say, $h=1/10$); and let $H^{\rm out}$ denote the remaining
integration area over the sphere, outside $H^{\rm in}$.
Correspondingly, we divide the integral $I^{\beta}_1$ into two
contributions, as $I^{\beta}_1=I^{\beta{\rm in}}_{1}+I^{\beta{\rm out}}_1$.
Now, since the integrand of $I^{\beta{\rm out}}_1$ contains no singularity
(the only singularity on the sphere occurs at $x=y=0$, which is
located in $H^{\rm in}$), in evaluating
$\lim_{\delta r\to 0^{\pm}}I^{\beta{\rm out}}_1(\delta r)$ we are
allowed to interchange the limit and integration:
\begin{equation}\label{VII-295}
\lim_{\delta r\to 0^{\pm}} I^{\beta{\rm out}}_1
=I^{\beta{\rm out}}_1(\delta r=0).
\end{equation}
Precisely as in the case of the integral $I^{\beta}_2$ considered above,
this contribution is then found to vanish for either $\beta=r$ or $\beta=x$.
We are thus left with
$\tilde F_{\pm}^{l\beta}=\lim_{\delta r\to 0^{\pm}}I^{\beta{\rm in}}_1$,
namely,
\begin{equation}\label{VII-300}
\tilde F_{\pm}^{l\beta}= \lim_{\delta r\to 0^{\pm}}
\int_{-h}^h \int_{-h}^h \delta x^{\beta}\epsilon_0^{-3} dxdy.
\end{equation}

We proceed by considering separately the two components $\beta=r$
and $\beta=x$. Let us begin with the $r$ component:
Re-scaling the integration variables as
$X\equiv x/\delta r$ and $Y\equiv y/\delta r$, we find
\begin{equation}\label{VII-310}
\tilde F_{\pm}^{lr}=\lim_{\delta r\to 0_{\pm}}
\int_{-h/\delta r}^{h/\delta r}\int_{-h/\delta r}^{h/\delta r}\,
[\tilde{\epsilon}_{\pm}(X,Y)]^{-3}dXdY,
\end{equation}
where
\begin{equation}\label{VII-320}
\tilde\epsilon_{\pm}\equiv \epsilon_0/\delta r=
\pm[g_{rr}^0+r_0^2(X^2+Y^2)+(u_r+u_x X)^2]^{1/2},
\end{equation}
and the $\pm$ sign refers to the sign of $\delta r$. Note that
$\tilde\epsilon_{\pm}$ (and hence the entire integrand) is independent
of $\delta r$, such that the $\delta r\to 0_{\pm}$ limit becomes trivial:
\begin{equation}\label{VII-330}
\tilde F_{\pm}^{lr}=\pm\int_{-\infty }^{\infty}\int_{-\infty }^{\infty}\,
[g_{rr}^0+r_0^2(X^2+Y^2)+(u_r+u_x X)^2]^{-3/2}dXdY\equiv \pm U.
\end{equation}
This is an elementary integral [see, e.g., Eq.\ (3.252-3) of \cite{GR},
in conjunction with Eq.\ (3.252-2) therein], yielding
\begin{equation}\label{VII-340}
U=(2\pi/r_0)[r_0^2 u_r^2+g_{rr}^0(r_0^2+u_x^2)]^{-1/2}.
\end{equation}
Note the relation
\begin{equation}\label{VII-345}
U=-\frac{2\pi f}{r_0^2 u_t},
\end{equation}
which stems directly from the ``radial'' geodesic equation of motion,
$(u^r)^2=u_t^2-(1-u_x^2/r_0^2)f$.

Consider next the case $\beta=x$. It is not possible to treat this case
the same as the case $\beta=r$, by changing the integration variables
to $X,Y$: doing so, the integrand becomes $X\tilde\epsilon_{\pm}^{-3}$,
and the double-integral does not strictly converge at infinity.
We therefore apply here a different method to evaluate the limit
$\delta r\rightarrow 0^{\pm}$ in Eq.\ (\ref{VII-300}).
First, we express $\tilde{F}_{\pm }^{lx}$ as
\begin{equation}\label{VII-350}
\tilde F_{\pm}^{lx}=
\lim_{\delta r\to 0^{\pm }}\int_{-h}^{h}\int_{-h}^{h}xS_{0}^{-3/2}dxdy,
\end{equation}
where, recall,
$S_0=\epsilon_0^2=r_0^2(x^2+y^2)+g_{rr}^0\delta r^2+(u_r\delta r+
u_x x)^2$ (with the limit $t\rightarrow t_{0}$ already taken).
Now, $S_0$ is quadratic in $\delta x^{\alpha}$, and its derivative
with respect to $x$ is a linear combination of both $x$ and $\delta r$.
One easily obtains the relation
\begin{equation}\label{VII-360}
x=\alpha S_{0,x}+\beta \delta r,
\end{equation}
where the coefficients $\alpha$ and $\beta$ are given by
\begin{equation}\label{VII-370}
\alpha=\frac{1}{2(r_0^2+u_x^2)},\quad\quad
\beta=-\frac{u_x u_r}{r_0^2+u_x^2}.
\end{equation}
Substituting $x$ from Eq.\ (\ref{VII-360}) in Eq.\ (\ref{VII-350}),
we express $\tilde F_{\pm}^{lx}$ as the sum of two integrals:
\begin{eqnarray}\label{VII-380}
\tilde F_{\pm}^{lx}=
\lim_{\delta r\to 0^{\pm}}\left[
\alpha \int S_{0,x}S_{0}^{-3/2}dxdy
+\beta \int \delta r\, \epsilon_0^{-3}dxdy\right]
\equiv \alpha I_{i} +\beta I_{ii}.
\end{eqnarray}
In what follows we show that $I_i$ vanishes, leaving us with only the
contribution from $I_{ii}$, which is just proportional to the
$r$-component $\tilde F_{\pm}^{lr}$ calculated above.

Considering first $I_i$, we carry out the trivial integration over $x$,
obtaining
\begin{eqnarray}\label{VII-390}
I_{i}=
\lim_{\delta r\to 0^{\pm}}
\int_{-h}^{h}\left[-2S_{0}^{-1/2}\right]^{x=h}_{x=-h}dy.
\end{eqnarray}
The integration over $y$ is then a standard one, but one do not need
to carry it out explicitly: Observing that the integrand is now a
regular function of $y$ and $\delta r$ throughout the entire range
of integration, we are allowed to interchange the limit and integration.
Noticing then
$S_0(\delta r=0,x=+h)=S_0(\delta r=0,x=-h)$, we immediately conclude
\begin{eqnarray}\label{VII-400}
I_{i}=0.
\end{eqnarray}
Consider next $I_{ii}$. Comparing with Eq.\ (\ref{VII-300}) (for the $r$
components) we find simply $I_{ii}=\tilde{F}_{\pm}^{lr}$, hence
\begin{eqnarray}\label{VII-410}
\tilde F_{\pm}^{lx}=\beta\tilde F_{\pm}^{lr}=\pm\beta U.
\end{eqnarray}

Having calculated all components of $\tilde F_{\pm}^{l\beta}$, we
may now construct $F_{\pm\alpha}^{l(A)}$ through Eq.\ (\ref{VII-230}).
We obtain
\begin{mathletters} \label{VII-420}
\begin{equation}\label{VII-420a}
F_{\pm t}^{l(A)}=\mp [L/(2\pi)]u_t(u_r+\beta u_x)U=
\pm \frac{L f u_r}{r_0^2+u_x^2},
\end{equation}
\begin{equation}\label{VII-420b}
F_{\pm r}^{l(A)}=
\mp [L/(2\pi)]\left[f^{-1}+u_r(u_r+\beta u_x)\right]U=
\pm \frac{L f^{-1} u_t}{r_0^2+u_x^2},
\end{equation}
\begin{equation}\label{VII-420c}
F_{\pm x}^{l(A)}=
\mp [L/(2\pi)]\left[\beta(r_0^2+u_x^2)+u_xu_r\right]U=0,
\end{equation}
\begin{equation}\label{VII-420d}
F_{\pm y}^{l(A)}=0,
\end{equation}
\end{mathletters}
where we have substituted for $U$ and $\beta$ from Eqs.\ (\ref{VII-345})
and (\ref{VII-370}), respectively.

Note the remarkable fact that the contribution $F_{\pm\alpha}^{l(A)}$ is
{\em precisely} proportional to $L$.

\section{Values of the regularization parameters}\label{SecVIII}

In conclusion of the calculation carried out in the previous section, we
have found that the $l$-mode direct force $F_{\pm\alpha}^{l}$ is composed
of only two contributions: one---completely described by
$F_{\pm\alpha}^{l(A)}$---is precisely proportional to $L$, and the
other---completely described by $F_{\pm\alpha}^{l(B)}$---is independent
of $L$. No other powers of $L$ are present. Recalling the definition of the
RP in Sec.\ \ref{SecII}, we then conclude that the term $F_{\pm\alpha}^{l(A)}$
contributes only to the parameter $A_{\alpha}$ and that the term
$F_{\pm\alpha}^{l(B)}$ contributes only to $B_{\alpha}$.
Recalling Eq.\ (\ref{VII-10}), we identify the RP as
\begin{equation}\label{VIII-10}
LA_{\pm\alpha} =q^2 F_{\pm\alpha}^{l(A)},\quad\quad
 B_{\alpha} =q^2 F_{\pm\alpha}^{l(B)},\quad\quad
C_{\alpha}=0.
\end{equation}
Furthermore, from Eq.\ (\ref{II-70}) we immediately get $D_{\alpha}=0$.
The explicit values of $A_{\pm\alpha}$ and $B_{\alpha}$ are then obtained
by substituting the expressions derived above for the quantities
$F_{\pm\alpha}^{l(A,B)}$---Eqs.\ (\ref{VII-130}),
(\ref{VII-220}), and (\ref{VII-420}).

To give a useful summary of the RP values thus obtained, we shall transform
the angular coordinates $x,y$ back to the {\em standard} $\theta,\varphi$
coordinates, in which the orbit is equatorial (i.e., confined to $\theta=
\pi /2$). The quantities $F_{r}^{\rm dir}(x)$ and $F_{t}^{\rm dir}(x)$ are
unaffected by this transformation, therefore the $r$ and $t$
components of all RP are unchanged. However, $F_{x}^{\rm dir}$ and
$F_{y}^{\rm dir}$ transform to $F_{\theta}^{\rm dir}$ and
$F_{\varphi}^{\rm dir}$ in a manner which is not completely trivial,
and we need to find the corresponding $\theta$ and $\varphi$ components of
the RP. Note that a-priori there is no guarantee that the RP
will transform like vectors at the evaluation point, because
the RP depend on $F_{\alpha}^{\rm dir}$ in the neighborhood of $z$;
and the transformation $(x,y)\to (\theta,\varphi)$ involves nontrivial
functions of the angular coordinates, which may affect the mode
decomposition. However, in Appendix \ref{AppC} we show that in fact all the
RP do transform (in this particular coordinate transformation) like
four-vectors at $z$. It is trivial to show that at the evaluation point
\begin{equation}\label{VIII-15}
x_{,\theta}=y_{,\varphi}=0\quad\quad
x_{,\varphi}=-y_{,\theta}=1
\end{equation}
[for concreteness we consider here the transformation described by a $\pi/2$
rotation about the horizontal axis $\varphi=0$, which takes $z$ from the
pole to the point $(\theta_{0},\varphi_{0})=(\pi/2,-\pi/2)$---see Fig.\
\ref{figure} and Appendix \ref{AppC} for more details. Note, however, that
the RP in the $\theta,\varphi$ coordinates do not depend on $\varphi_{0}$,
due to the symmetry of rotations in $\varphi$.] Therefore,
$B_{\varphi}=B_{x}$ and $B_{\theta}=-B_{y}$ (and the
same for $A_{\varphi},A_{\theta}$). Note that $u_{x}(z)$ now becomes
$u_{\varphi}\equiv \mathcal{L}$, the conserved azimuthal angular momentum.
In the standard $\theta,\varphi $ coordinates, the RP are then given by
$A_{\theta}=B_{\theta}=0$,
\begin{mathletters} \label{RP}
\begin{equation}\label{A}
A_{\pm r}=\mp\frac{q^2}{r_0^2}\,\frac{\cal E}{fV},\quad
A_{\pm t}=\pm\frac{q^2}{r_0^2}\,\frac{\dot{r}}{V},\quad
A_{\varphi}=0,
\end{equation}
\begin{equation}\label{Br}
B_r=\frac{q^2}{r_0^2}\,
\frac{(\dot{r}^2-2{\cal E}^2)\hat{K}(w)+(\dot{r}^2+E^2)\hat{E}(w)}
{\pi f V^{3/2}},
\end{equation}
\begin{equation}\label{Bt}
B_t=\frac{q^2}{r_0^2}\,\frac{{\cal E}\dot{r}[\hat{K}(w)-2\hat{E}(w)]}
{\pi V^{3/2}},
\end{equation}
\begin{equation}\label{Bphi}
B_{\varphi}=\frac{q^2}{r_0}\,\frac{\dot{r}[\hat{K}(w)-\hat{E}(w)]}
{\pi ({\cal L}/r_0)V^{1/2}},
\end{equation}
\begin{equation}\label{CD}
C_{\alpha}=D_{\alpha}=0,
\end{equation}
\end{mathletters}
where
\begin{equation}\label{CD2}
w\equiv \frac{{\cal L}^2}{{\cal L}^2+r_0^2}, \quad\quad
V\equiv 1+{\cal L}^2/r_0^2,
\end{equation}
$f\equiv(1-2M/r_0)$, and $\dot{r}^2\equiv(u^r)^2={\cal E}^2-fV$.
Recall ${\cal E}=-u_t$ and ${\cal L}=u_{\varphi}$ are the (conserved)
specific energy and angular momentum parameters.

We comment that the parameter $A_{\pm\alpha}$ is normal to the
four-velocity: $A_{\pm\alpha}u^{\alpha}=0$.
However, the parameter $B_{\alpha}$ (as the self force itself in our model)
is, in general, {\em not} normal to $u^{\alpha}$: An explicit calculation
yields
\begin{equation}\label{VIII-20}
B_{\alpha}u^{\alpha}=-2\,\frac{q^2}{r_0^2}\,
\frac{\dot{r}\hat{E}(w)}{\pi V^{1/2}}.
\end{equation}

Finally, we give here the RP values for the special case of a radial
geodesic, i.e., ${\cal L}=0$: Noting, in this case, $w=0$, $V=1$, and
$\dot{r}^{2}={\cal E}^{2}-f$, and recalling $\hat{K}(0)=\hat{E}(0)=\pi /2$,
the nonvanishing components in Eqs.\ (\ref{RP}) reduce to
\begin{mathletters} \label{RP-radial}
\begin{equation}\label{A-radial}
A^{\rm radial}_{\pm r}=\mp\frac{q^2}{r_0^2}({\cal E}/f),\quad\quad
A^{\rm radial}_{\pm t}=\pm\frac{q^2}{r_0^2}\,\dot{r},
\end{equation}
\begin{equation}\label{B-radial}
B^{\rm radial}_r=\frac{q^2}{2r_0^2}\,f^{-1}\left({\cal
E}^2-2f\right),\quad\quad
B^{\rm radial}_t=-\frac{q^2}{2r_0^2}\,{\cal E}\dot{r}.
\end{equation}
\end{mathletters}
[The vanishing of $B_{\varphi }^{\rm{radial}}$ is obvious from symmetry
considerations. Note that $B_{\varphi }$ vanishes at the limit $w\to 0$
despite the factor $\cal{L}$ in the denominator, because
$\hat{K}(w)-\hat{E}(w)=O(w)=O({\cal L}^{2})$.] These values are in agreement
with the ones derived in \cite{MSRS-scalar} using the $l$ mode Green's
function expansion method.


\section{Concluding remarks} \label{SecIX}

The mode-sum scheme described by Eq.\ (\ref{II-80}), with the explicit
RP values calculated in this paper, Eqs.\ (\ref{RP}), provide one with
a practical means---yet one based on a physically well-established
regularization scheme---for calculating the scalar self-force for
any geodesic orbit around a Schwarzschild black hole. Recall that
the full modes $F_{\alpha}^{{\rm (full)}l}$ needed for fully
implementing this mode-sum scheme, are to be obtained from
the $l$-modes of the scalar field $\Phi$, which, in turn, are to
be calculated using standard numerical techniques.

The RP values derived here, Eqs.\ (\ref{RP}), were obtained independently
by MNS \cite{MNS} using a different approach. In their analysis, MNS
decomposed the direct field using the ``standard'' $\theta,\varphi$
coordinates (in which the motion is equatorial), in which case the
$m\neq 0$ modes contribute as well. MNS then derived an analytic expression
for the contribution of each $l,m$ mode of the direct force, expanded in
powers of $M/r$. By explicitly summing up this expansion (and summing over
$m$), MNS were able to recover all RP values.

The RP values (\ref{RP}) reduce, in the special cases of radial
motion (${\cal L}=0$) or circular motion ($\dot{r}=0$), to the values
derived previously \cite{MSRS-scalar} using a completely independent
analytic approach (namely, by locally analyzing the $l$-mode Green's function,
as we briefly mention in the introduction). In these cases, the values of the
parameters $A_{\alpha}$, $B_{\alpha}$, and $C_{\alpha}$ have also been
confirmed numerically, by calculating the full-force modes
\cite{implementation}.

The RP calculation method presented in this paper is directly applicable
to the more realistic case of the {\em gravitational} self-force acting on
a mass particle, as well as to the case of the {\em electromagnetic}
self-force acting on an electrically-charged particle. Both cases shall
be treated in an accompanying paper \cite{paperII}, where we obtain the
gravitational and electromagnetic RP for general orbits in Schwarzschild
spacetime (the results in the gravitational case were provided in
\cite{Letter}). The extension of our scalar field analysis to the
gravitational and electromagnetic cases involves several complexities
which require a special care. In particular, one has to tackle the technical
issue of extending the four-velocity vector off the worldline \cite{paperII}.
A more fundamental issue concerns the gauge dependence of the gravitational
self-force \cite{gauge}.



\section*{acknowledgements}
We are grateful to Lior Burko, Yasushi Mino, Hiroyuki Nakano, and
Misao Sasaki for interesting discussions and stimulating interaction.
L.B.\ was supported by a Marie Curie Fellowship of the European Community
program IHP-MCIF-99-1 under contract number HPMF-CT-2000-00851.

\appendix

\section{Derivation of $S_0$ and $S_1$} \label{AppA}

In this appendix we calculate the two leading terms in the expansion of
$S\equiv S(x,z)$ [the square of the geodesic distance from the point $x$ to
the geodesic $z(\tau)$] in powers of $\delta x^{\mu}\equiv x^{\mu}-z^{\mu}$.
This expansion takes the form
\begin{equation}\label{A-10}
S=S_{0}+S_{1}+S_{2}+\cdots,
\end{equation}
in which the term $S_{n}$ is of homogeneous order $\delta x^{n+2}$, and we
wish to calculate $S_{0}$ and $S_{1}$.

In flat space, using Cartesian coordinates $y^{\alpha}$ (with $y^{\alpha}=0$
at $z$), we obviously have $S=(\eta_{\alpha\beta}+u_{\alpha}u_{\beta})
y^{\alpha}y^{\beta}\equiv S_{0}$, where $\eta_{\alpha\beta}$ is the flat
space metric. In curved space (or in curvilinear coordinates), each of the
terms $S_{n}$ (like $S$ itself) will be a certain function of
$g_{\alpha\beta}$ and its derivatives. From simple dimensionality
considerations, it is clear that $S_{0}$ may not include any derivatives of
$g_{\alpha\beta}$, and that $S_{1}$ may include only first-order derivatives
of the latter (in addition to $g_{\alpha\beta}$ itself) .

Let $y^{\alpha}$ be locally-Cartesian coordinates at the evaluation point
$z$, with $y^{\alpha}=0$ at $z$. Namely, at $x=z$, the metric functions in
the coordinates $y^{\alpha}$ are just $\eta_{\mu\nu}$, and their
first-order derivatives vanish. Since no second or higher-order derivatives
appear in $S$ up to the desired order, we must have
\begin{equation} \label{A-20}
S=(\eta_{\alpha\beta}+u'_{\alpha}u'_{\beta})y^{\alpha}y^{\beta}
+O(y^{4}),
\end{equation}
where a prime denotes vectorial components in the $y^{\alpha}$ coordinate
system. We now transform from $y^{\alpha}$ back to our original coordinates
$x^{\mu}$. Recall that $S$ is a bi-scalar, and is hence invariant under
this transformation. Writing the Taylor expansion of
$y^{\alpha }(\delta x^{\mu })$,
\begin{equation}\label{A-30}
y^{\alpha}=\frac{\partial y^{\alpha}}{\partial x^{\lambda}}
\delta x^{\lambda}+\frac{1}{2}\frac{\partial^{2}y^{\alpha}}{\partial x^{\mu}
\partial x^{\nu}}\delta x^{\mu}\delta x^{\nu}+O(\delta x^{3})
\end{equation}
(in which all coefficients are evaluated at $z$), and substituting it in the
right-hand side of Eq. (\ref{A-20}), we find
\begin{eqnarray*}
S &=&\left[(\eta_{\alpha\beta}+u'_{\alpha}u'_{\beta})
\frac{\partial y^{\alpha}}{\partial x^{\lambda}}\frac{\partial y^{\beta}}
{\partial x^{\epsilon}}\right] \delta x^{\lambda}\delta x^{\epsilon}+
\\
&&\left[(\eta_{\alpha\beta}+u'_{\alpha}u'_{\beta})
\frac{\partial y^{\alpha}}{\partial x^{\lambda}}\frac{\partial^{2}y^{\beta}}
{\partial x^{\mu}\partial x^{\nu}}\right]\delta x^{\lambda}\delta x^{\mu }
\delta x^{\nu}+O(\delta x^{4}).
\end{eqnarray*}
Comparing this to Eq. (\ref{A-10}), we identify the first and second terms
on the right-hand side with $S_{0}$ and $S_{1}$, respectively. Using the
obvious tensorial transformation rule, we find
\begin{equation} \label{A-40}
S_{0}=(g_{\lambda\epsilon}+u_{\lambda}u_{\epsilon})\delta x^{\lambda}
\delta x^{\varepsilon}.
\end{equation}
To calculate $S_{1}$ we need the second-order transformation coefficients,
which are given by
\begin{eqnarray*}
\frac{\partial^{2}y^{\beta}}{\partial x^{\mu}\partial x^{\nu}}=
\Gamma_{\mu\nu}^{\epsilon}\frac{\partial y^{\beta}}{\partial x^{\epsilon}}
\end{eqnarray*}
[see, e.g., Eq. (3.2.11) of \cite{W}]. Therefore,
\begin{eqnarray*}
S_{1} &=&\left[(\eta_{\alpha\beta}+u'_{\alpha}u'_{\beta})
\frac{\partial y^{\alpha}}{\partial x^{\lambda}}\frac{\partial y^{\beta}}
{\partial x^{\epsilon}}\Gamma_{\mu\nu}^{\epsilon}\right] \delta x^{\lambda}
\delta x^{\mu}\delta x^{\nu} \\
&=&(g_{\lambda\epsilon}+u_{\lambda}u_{\epsilon})\Gamma_{\mu\nu}^{\epsilon}
\delta x^{\lambda}\delta x^{\mu}\delta x^{\nu}.
\end{eqnarray*}
Recalling that
\begin{eqnarray*}
g_{\lambda\epsilon}\Gamma_{\mu\nu}^{\epsilon}\delta x^{\lambda}
\delta x^{\mu}\delta x^{\nu}=\frac{1}{2}g_{\mu\nu,\lambda}\delta x^{\lambda}
\delta x^{\mu}\delta x^{\nu},
\end{eqnarray*}
we finally obtain
\begin{equation} \label{A-50}
S_1=\left(u_{\lambda}u_{\gamma}\Gamma^{\lambda0}_{\alpha\beta}+
g^0_{\alpha\beta,\gamma}/2 \right)
\delta x^{\alpha}\delta x^{\beta}\delta x^{\gamma}.
\end{equation}

\section{Interchangeability of the $\lowercase{r}\to \lowercase{r}_0$
limit and the Legendre integral} \label{AppB}

In this appendix we explore the interchangeability of the limit and
integration in Eq.\ (\ref{VII-20})---an issue crucial for the calculation
carried out in Sec.\ \ref{SecVII}. For convenience, let us write
$F_{\pm\alpha}^{(W)l}=[L/(2\pi)]\hat F_{\pm\alpha}^{(W)l}$, where
\begin{equation}\label{B-10}
\hat F_{\pm\alpha}^{(W)l} \equiv \lim_{\delta r\to 0^{\pm}}
\int_{0}^{\pi}d\theta \int_{0}^{2\pi}d\varphi \,
F_{\alpha}^{(W)}(r,t_0,\theta,\varphi)P_l(\cos\theta)\sin\theta.
\end{equation}
Here, recall, $W$ stands for $A$, $B$, or $C$, with
\begin{equation}\label{B-20}
F_{\alpha}^{(A)}\equiv \epsilon_0^{-3}P_{\alpha}^{(1)},\quad\quad
F_{\alpha}^{(B)}\equiv \epsilon_0^{-5}P_{\alpha}^{(4)},\quad\quad
F_{\alpha}^{(C)}\equiv \epsilon_0^{-7}P_{\alpha}^{(7)},
\end{equation}
where $\epsilon_0$ is given explicitly in Eq.\ (\ref{VII-250}), and
$P_{\alpha}^{(n)}$ represents a polynomial of homogeneous
order $n$ in $\delta x^{\mu}\equiv x^{\mu}-z^{\mu}$.
We shall show that interchanging the limit and integration in
Eq.\ (\ref{B-10}) is valid for $W=B,C$, and explain why our proof
fails in the case $W=A$.

We begin by considering the case $W=B$.
Let $R\equiv |\delta r|^{s}$ for some $0<s<1$. Consider the range of small
$\delta r$ (such that $R<1$). We split the $\theta$-integral in Eq.\
(\ref{B-10}) into three domains: (i) $\theta<R$, (ii) $R<\theta <1$, and
(iii) $1<\theta<\pi$. Correspondingly, the above double-integral can be
expressed as ${\cal I}_{i}+{\cal I}_{ii}+{\cal I}_{iii}$, hence
\begin{equation}\label{B-30}
\hat{F}_{\pm\alpha}^{(B)l}=
\lim_{\delta r\to 0^{\pm}} {\cal I}_{i}(\delta r)+
\lim_{\delta r\to 0^{\pm}} {\cal I}_{ii}(\delta r)+
\lim_{\delta r\to 0_{\pm}} {\cal I}_{iii}(\delta r).
\end{equation}

Consider first the internal integral ${\cal I}_{i}$:
\begin{equation}\label{B-40}
{\cal I}_{i}(\delta r)=\int_0^R d\theta \int_0^{2\pi}d\varphi\,
F_{\alpha}^{(B)}(\delta r,\theta,\varphi)\,P^{l}(\cos\theta)\sin\theta.
\end{equation}
Since at the relevant limit $R\to 0$, we have $\theta\ll 1$
throughout the range of integration, and we may approximate this integral
[using $\sin\theta d\theta\simeq \rho d\rho$, as well as
$P^{l}(\cos\theta )\simeq 1$ and $\rho (R)\simeq R$] as
\begin{equation}\label{B-50}
{\cal I}_{i}(\delta r)\simeq \int_0^R\rho d\rho \int_0^{2\pi}d\varphi\,
F_{\alpha}^{(B)}(\delta r,\rho,\varphi).
\end{equation}
Now, $F_{\alpha}^{(B)}=\epsilon_0^{-5}P_{\alpha}^{(4)}$. We may bound
$\epsilon_0$ and $P_{\alpha}^{(4)}$ as $\epsilon_0>c_1|\delta r|$ and
$|P_{\alpha}^{(4)}|<c_2\rho^4$, where hereafter $c_{n}$ are some positive
constants. Consequently, $F_{\alpha}^{(B)}$ can be bounded as
$|F_{\alpha}^{(B)}|<c_{3}|\delta r|^{-5}\rho^4<c_3|\delta r|^{-5}R^4$.
Since the ''integration area'' in Eq.\ (\ref{B-50}) is $\pi R^2$,
we then obtain the upper bound
\begin{equation}\label{B-60}
|{\cal I}_{i}|<\pi c_3|\delta r|^{-5}R^6=c_4|\delta r|^{6s-5}.
\end{equation}
Taking, e.g., $s=0.9$, we find that $|{\cal I}_{i}|<c_4|\delta r|^{0.4}$ and
hence it vanishes at the limit $\delta r\to 0^{\pm}$:
\begin{equation}\label{B-70}
\lim_{\delta r\to 0_{\pm}}{\cal I}_{i}(\delta r)=0.
\end{equation}

Consider next ${\cal I}_{ii}$, which is defined by the double-integral
\begin{equation}\label{B-80}
{\cal I}_{ii}(\delta r)=\int_R^1 d\theta \int_0^{2\pi}d\varphi\,
F_{\alpha}^{(B)}(\delta r,\theta,\varphi)\,P^l(\cos\theta)\sin\theta.
\end{equation}
Let us introduce the quantity $\Delta F_{\alpha}^{(B)}\equiv F_{\alpha}^{(B)}
(\delta r)-F_{\alpha}^{(B)}(\delta r=0)$, and the corresponding integral
\begin{equation}\label{B-100}
\Delta{\cal I}_{ii}(\delta r)\equiv\int_R^1 d\theta \int_0^{2\pi}d\varphi\,
\Delta F_{\alpha}^{(B)}(\delta r,\theta,\varphi)\,P^l(\cos\theta)\sin\theta,
\end{equation}
such that
\begin{equation}\label{B-110}
{\cal I}_{ii}(\delta r)=\Delta{\cal I}_{ii}(\delta r)+\int_R^1 d\theta
\int_0^{2\pi}d\varphi\, F_{\alpha}^{(B)}(\delta r=0,\theta,\varphi)
P^l(\cos\theta)\sin\theta.
\end{equation}
Since $s<1$, in the entire range (ii) we have $|\delta r|\ll R\leq\theta$
(for $|\delta r|\ll 1$), and thus $|\delta r|\ll \rho$.
We may approximate $\Delta F_{\alpha}^{(B)}$ at $|\delta r|\ll 1$ as
\begin{equation}\label{B-90}
\Delta F_{\alpha}^{(B)}\simeq
\left. d(\epsilon_0^{-5}P_{\alpha}^{(4)})/d(\delta r)
\right|_{\delta r=0}\times \delta r.
\end{equation}
From Eq.\ (\ref{VII-250}) (and recalling $|x|\leq\rho$) we observe that,
at $\delta r=0$, $\epsilon_0^{-1}$ and $d(\epsilon_0^{-1})/d(\delta r)$ are
bounded from above by $\propto\rho^{-1}$ and $\propto{\rm const}$,
respectively. At $\delta r=0$ we may also upper bound $P_{\alpha}^{(4)}$
and $dP_{\alpha}^{(4)}/d(\delta r)$ by $\propto\rho^4$ and $\propto\rho^3$,
respectively. Applying these bounds in Eq.\ (\ref{B-90}) we then obtain
\begin{equation}\label{B-95}
|\Delta F_{\alpha}^{(B)}|<c_5|\delta r|\rho^{-2}.
\end{equation}
[Note that in Eq.\ (\ref{B-90}) we have assumed that either $S$ or
$P_{\alpha}^{(4)}$ (or both) include terms linear in $\delta r$.
In special cases where only terms quadratic in $\delta r$ are present
in both quantities, we shall instead arrive at the bound
$|\Delta F_{\alpha}^{(B)}|<c_6\delta r^2\rho^{-3}$.
Since $|\delta r|/\rho\ll 1$, this is even smaller than the bound
(\ref{B-95}), and so the entire derivation below remains valid.]

Using Eq.\ (\ref{B-95}) and recalling $|P^l(\cos\theta)|\leq 1$, we can
now bound $\Delta I_{ii}(\delta r)$ by
\begin{equation}\label{B-120}
|\Delta {\cal I}_{ii}(\delta r)| <c_{5}\int_R^1 d\theta
\int_{0}^{2\pi}d\varphi \sin\theta\, |\delta r| \rho^{-2}=
2\pi c_{5}\int_R^1 d\theta \sin\theta\, |\delta r| \rho^{-2}.
\end{equation}
Since $\sin\theta/\rho$ and $d\theta/d\rho $ are bounded in domain (ii),
the last integral can now be bounded as
\begin{equation}\label{B-130}
|\Delta {\cal I}_{ii}(\delta r)| <
c_6|\delta r|\int_{\rho(R)}^{\rho(1)}\rho^{-1}d\rho.
\end{equation}
[We point out here that in the way we have chosen $\rho(\theta)$ in Sec.\
\ref{SecVII}---see Eq.\ (\ref{VII-160})---$d\theta/d\rho$
is unbounded at $\theta\to\pi$. It is this divergence that forced us to
terminate domain (ii) at $\theta=1$, and to introduce domain (iii).]
Upon integration we find
\begin{equation}\label{B-140}
|\Delta{\cal I}_{ii}(\delta r)|<c_6|\delta r|
\left[\ln\rho (1)-\ln\rho (R)\right].
\end{equation}
For small $R$ we have $\rho(R)\simeq R=|\delta r|^s$; therefore
\begin{equation}\label{B-150}
|\Delta{\cal I}_{ii}(\delta r)|< c_6\ln\rho(1)|\delta r|
-sc_6|\delta r|\ln|\delta r|.
\end{equation}
Clearly, both terms vanish as $\delta r\to 0^{\pm}$, hence
\begin{equation}\label{B-160}
\lim_{\delta r\to 0^{\pm}}\Delta{\cal I}_{ii}(\delta r)=0.
\end{equation}
From Eq.\ (\ref{B-110}) we then have
\begin{eqnarray}\label{B-170}
\lim_{\delta r\to 0^{\pm}}{\cal I}_{ii}(\delta r) &=&
\lim_{\delta r\to 0^{\pm}}\int_{R(\delta r)}^{1}d\theta \int_0^{2\pi}
d\varphi\, F_{\alpha}^{(B)}(\delta r=0,\theta,\varphi)\,P^l(\cos\theta)
\sin\theta
\nonumber\\
&=&\int_0^{1}d\theta\int_0^{2\pi}d\varphi\, F_{\alpha}^{(B)}
(\delta r=0,\theta,\varphi)\,P^l(\cos\theta)\sin\theta.
\end{eqnarray}

Finally, consider the third contribution, ${\cal I}_{iii}$, defined by
\begin{equation}\label{B-180}
{\cal I}_{iii}(\delta r)=\int_1^{\pi}d\theta \int_0^{2\pi}d\varphi\,
F_{\alpha}^{(B)}(\delta r,\theta,\varphi)\,P^l(\cos\theta)\sin\theta.
\end{equation}
In full analogy with the analysis above, we define
\begin{equation}\label{B-190}
\Delta{\cal I}_{iii}(\delta r)=\int_1^{\pi}d\theta\int_0^{2\pi}d\varphi\,
\Delta F_{\alpha}^{(B)}(\delta r,\theta,\varphi)\,P^l(\cos\theta)\sin\theta.
\end{equation}
The above bound, $|\Delta F_{\alpha}^{(B)}|<c_5|\delta r|\rho^{-2}$,
is valid in range (iii) too. Since in this range $\rho$ is bounded from
below, we may now write $|\Delta F_{\alpha}^{(B)}|<c_{7}|\delta r|$;
and since $P^{l}(\cos\theta)$ and $\sin\theta$ are both bounded by
unity, the entire integrand is then bounded by $c_7|\delta r|$:
\begin{equation}\label{B-200}
|\Delta{\cal I}_{iii}(\delta r)|< c_{7}|\delta r|\int_{1}^{\pi}d\theta
\int_{0}^{2\pi}d\varphi = 2\pi^{2}c_{7}|\delta r|.
\end{equation}
Again, this quantity vanishes at the limit $\delta r\to 0$; hence,
by the same considerations used above for range (ii)
[see the chain of equations (\ref{B-170})], we obtain
\begin{equation}\label{B-210}
\lim_{\delta r\to 0_{\pm}}{\cal I}_{iii}(\delta r)=\int_{1}^{\pi}d\theta
\int_{0}^{2\pi}d\varphi\, F_{\alpha}^{(B)}(\delta r=0,\theta,\varphi)\,
P^{l}(\cos\theta)\sin\theta.
\end{equation}

Substituting Eqs.\ (\ref{B-70}), (\ref{B-170}), and
(\ref{B-210}) in Eq.\ (\ref{B-30}), we obtain
\begin{equation}\label{B-220}
\hat{F}_{\alpha}^{l(B)}=
\int_0^{\pi}d\theta\int_0^{2\pi}d\varphi\, F_{\alpha}^{(B)}(\delta r=0,
\theta ,\varphi )\,P^l(\cos\theta)\sin\theta.
\end{equation}
Namely, in the calculation of $\hat F_{\pm\alpha}^{l(B)}$---and thus
also $F_{\pm\alpha}^{l(B)}$---we are allowed to
interchange the limit $\delta r\to 0$ and the integration.
Note that since $\hat F_{\alpha}^{(B)}$ admits a well-defined limit
at $\delta r\to 0$ (except at $\theta=0$---which, however, was shown
not to affect the integral), we have omitted its $\pm$ label.

The same proof can immediately be applied to $\hat{F}_{\pm \alpha }^{l(C)}$.
Evaluating ${\cal I}_{i}$, we find this time that $|F_{\alpha}^{(C)}|$ is
bounded from above by $c_8|\delta r|^{-7}\rho^7<c_8|\delta r|^{-7}R^7$,
hence (taking again $s=0.9$)
\begin{equation}\label{B-230}
|{\cal I}_{i}|<c_9|\delta r|^{-7}R^{9}=c_9|\delta r|^{9s-7}
=c_9|\delta r|^{1.1}\to 0
\end{equation}
as $\delta r\to 0$.
Evaluating next $\Delta {\cal I}_{ii}$, we obtain this time
[in analogy with Eq.\ (\ref{B-95})]
$|\Delta F_{\alpha}^{(C)}|<c_{10}|\delta r|/\rho$,
and hence
\begin{equation}\label{B-250}
|\Delta {\cal I}_{ii}(\delta r)|<c_{11}|\delta r|\int_{\rho (R)}^{\rho (1)}
\rho^0d\rho < c_{12}|\delta r|,
\end{equation}
which again vanishes as $\delta r\to 0$.
The calculation of $\Delta {\cal I}_{iii}$ proceeds exactly as for $W=B$
(the only difference is that now $\Delta F_{\alpha}^{(C)}\propto |\delta r|/\rho$
instead of $|\delta r|/\rho ^{2}$, but this does not affect the above
evaluation of $\Delta {\cal I}_{iii}$ in any way).
Again we find $|\Delta{\cal I}_{iii}(\delta r)|<2\pi^2c_{13}|\delta r|$,
which vanishes at the limit $\delta r\to 0$.
We conclude that the limit and integration may be interchanged for
$W=C$ as well:
\begin{equation}\label{B-260}
\hat F_{\alpha}^{l(C)}=\int_0^{\pi }d\theta\int_0^{2\pi}d\varphi\,
F_{\alpha}^{(C)}(\delta r=0,\theta,\varphi)\,P^l(\cos\theta)\sin\theta.
\end{equation}

Finally, it is instructive to see how the above type of arguments fail for
$W=A$. Since $F_{\alpha}^{(A)}=\epsilon_0^{-3}P_{\alpha}^{(1)}$, in
evaluating ${\cal I}_{i}$ one obtains
$|{\cal I}_{i}|<c_{14}|\delta r|^{-3}R^{3}=c_{14}|\delta r|^{3(s-1)}$.
Then, evaluating $\Delta{\cal I}_{ii}$, one obtains
\begin{equation}\label{B-270}
|\Delta{\cal I}_{ii}(\delta r)|<c_{15}|\delta r|\int_{\rho(R)}^{\rho (1)}
\rho^{-2}d\rho,
\end{equation}
which at the limit of small $R$ yields
$|\Delta{\cal I}_{ii}(\delta r)|<c_{15}|\delta r|/R=c_{15}|\delta r|^{1-s}$.
Obviously, for any $s\leq 1$ the bound for ${\cal I}_{i}$ will fail to vanish
as $\delta r\to 0$, and for any $s\geq 1$ the bound for $\Delta{\cal I}_{ii}$
will fail to vanish at this limit. [Note also that in the case $s\geq 1$ the
inequality $|\delta r|\ll \rho $, used above in evaluating
$\Delta{\cal I}_{ii}$, is no longer valid throughout range (ii)].
In fact, it becomes evident from the explicit calculation in Sec.\ \ref{SecVII},
that for $W=A$ the limit $\delta r\to 0$ {\em cannot} be interchanged with the
integration.

\section{Transforming to equatorial orbit}
\label{AppC}

In Sec.\ \ref{SecVII} we analyzed the multipole decomposition of the direct
force in a Schwarzschild coordinate system in which the particle is
momentarily at the pole. We then transformed to locally
Cartesian angular coordinates $x,y$ and calculated the RP in the system
$x^{\mu}_{\rm po}\equiv (t,r,x,y)$. Usually (e.g., in numerical calculations)
one adopts a more natural pair of angular coordinates, in which the particle's
orbit is confined to the equatorial plain ($\theta=\pi/2$)---throughout this
appendix we shall denote this system by $x^{\mu}_{\rm eq}\equiv
(t,r,\theta,\varphi)$. The goal of this Appendix is, given the RP values in
the system $x^{\mu}_{\rm po}$, to obtain the corresponding values in the
system $x^{\mu}_{\rm eq}$.

\begin{figure}[thb]
\input{epsf}
\centerline{\epsfysize 6cm \epsfbox{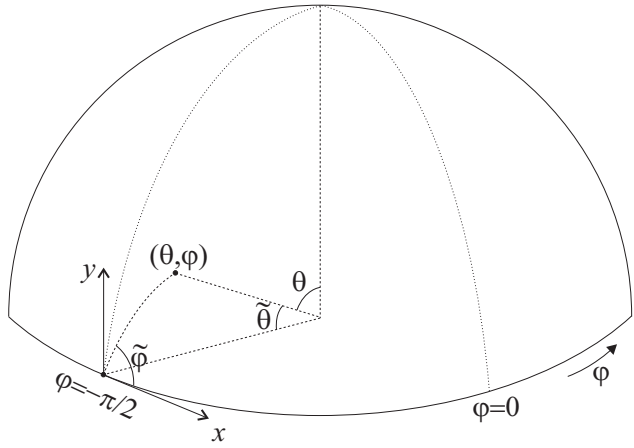}}
\caption{\protect\footnotesize
A Sketch showing the various quantities involved in constructing the
coordinate transformation $(x,y)\to(\theta,\varphi)$.
Shown is the ``northern'' hemisphere $t=r={\rm const}$, $\theta\geq 0$.
The particle moves along an equatorial orbit, and is momentarily located at
$(\theta,\varphi)=(\pi/2,-\pi/2)$. $x,y$ is a pair of ``locally Cartesian
angular coordinates'' at the particle's location, as described in the text.}
\label{figure}
\end{figure}

In the system $x^{\mu}_{\rm eq}$ we have $u^{\theta}=0$ and $u_{\varphi}=
\cal L$, where, recall, $\cal L$ is the conserved specific angular momentum.
Let the particle be momentarily at $(\theta,\varphi)=(\theta_0,\varphi_0)=
(\pi/2,-\pi/2)$.\footnote{
We consider here a specific value of $\varphi_0$ in order to simplify the
following expressions, and to make the correspondence between the
$\tilde x,\tilde y$ and $x,y$ coordinates (see below) easily apparent.
Note, however, that our final result---the RP values in the system
$x^{\mu}_{\rm eq}$---do not depend on the choice of $\varphi_0$, as the
$l$-mode decomposition of the direct force in that system is invariant
under rotations about the polar axis.}
Consider the set of spherical coordinates $\tilde\theta,\tilde\varphi$,
defined such that the particle is located at their pole, $\tilde\theta=0$,
and $\tilde\varphi=0,\pi$ coincides with $\theta=\pi/2$---see Fig.\
\ref{figure}. (These are, in fact, the same spherical coordinates used
throughout the paper; here we merely use a different notation, as the
symbols $\theta,\varphi$ are reserved for the angular coordinates of
the $x^{\mu}_{\rm eq}$ system.)
The ``locally Cartesian'' coordinates at $z$ are then given by
[see Eq.\ (\ref{VI-10})]
\begin{eqnarray}\label{C-10}
x&=&\rho(\tilde\theta)\cos\tilde\varphi=
2\sin(\tilde\theta/2)\cos\tilde\varphi,
\nonumber\\
y&=&\rho(\tilde\theta)\sin\tilde\varphi
=2\sin(\tilde\theta/2)\sin\tilde\varphi.
\end{eqnarray}
Relating the spherical coordinates $\tilde\theta,\tilde\varphi$ to the
standard pair $\theta,\varphi$ is a straightforward geometrical problem,
and one finds
\begin{equation}\label{C-20}
\cos\tilde\theta=-\sin\theta \sin\varphi,
\quad\quad
\cot\tilde\varphi=\tan\theta \cos\varphi.
\end{equation}
This allows us to express $x,y$ directly as functions
of $\theta,\varphi$. We obtain
\begin{eqnarray}\label{C-30}
x(\theta,\varphi)&=&
2^{1/2}\frac{\sin\theta\cos\varphi}{\sqrt{1-\sin\theta\sin\varphi}},
\nonumber\\
y(\theta,\varphi)&=&
2^{1/2}\frac{\cos\theta}{\sqrt{1-\sin\theta\sin\varphi}},
\end{eqnarray}
which describes explicitly the transformation between $x^{\mu}_{\rm eq}$
and $x^{\mu}_{\rm po}$. Note that this transformation is regular on the
entire sphere [except at the point $(\theta,\varphi)=(\pi/2,\pi/2$), which,
however, is irrelevant for our analysis].

Given the above transformation rule,
the desired components $F^{\rm dir}_{\theta}$ and $F^{\rm dir}_{\varphi}$
are constructed as
\begin{eqnarray}\label{C-35}
F^{\rm dir}_{\theta}&=&
x_{,\theta}F^{\rm dir}_{x}+ y_{,\theta}F^{\rm dir}_{y},    \nonumber\\
F^{\rm dir}_{\varphi}&=& x_{,\varphi}F^{\rm dir}_{x}+y_{,\varphi}F^{\rm dir}_{y}.
\end{eqnarray}
It is now useful to consider the Taylor expansion of the various partial
derivatives about the particle's location: Introducing
$\Delta\theta\equiv\theta-\pi/2$ and
$\Delta\varphi\equiv\varphi-(-\pi/2)$, we find
\begin{mathletters}\label{C-40}
\begin{equation}\label{C-40a}
x_{,\theta}=-\frac{3}{4}\Delta\theta\Delta\varphi
+O(\delta x^4),
\end{equation}
\begin{equation}\label{C-40b}
x_{,\varphi}=1-\frac{1}{8}\Delta\varphi^2
-\frac{3}{8}\Delta\theta^2+O(\delta x^4),
\end{equation}
\begin{equation}\label{C-40c}
y_{,\theta}=-1-\frac{1}{8}\Delta\varphi^2
+\frac{1}{8}\Delta\theta^2+O(\delta x^4),
\end{equation}
\begin{equation}\label{C-40d}
y_{,\varphi}=-\frac{1}{4}\Delta\theta\Delta\varphi
+O(\delta x^4),
\end{equation}
\end{mathletters}
where $O(\delta x^4)$ represents corrections of fourth order in $\Delta\theta$
and $\Delta\varphi$. Substituting these exapnsions in Eqs.\ (\ref{C-35}), we
obtain, near the particle's location,
\begin{equation}\label{C-50}
F^{\rm dir}_{\theta}\simeq -F^{\rm dir}_y,
\quad\quad F^{\rm dir}_{\varphi}\simeq F^{\rm dir}_x,
\end{equation}
where corrections are due to terms of the form $\epsilon_0^{-7}P^{(7)}_{\alpha}$
(recall the notation introduced in Sec.\ \ref{SecIV}), and higher-order terms
that vanish at $x\to z$. From the analysis of Sec.\ \ref{SecVII} it is clear
that such correction terms do not contribute to neither $F_{\alpha}^{(A)}$
nor $F_{\alpha}^{(B)}$, and their contribution to $F_{\alpha}^{(C)}$
vanishes in the multipole decomposition. Hence, none of the RP will
be affected by omitting these correction terms and replacing the
approximation in Eq.\ (\ref{C-50}) with an exact equality.
Consequently, we find that the RP transform under
$x^{\mu}_{\rm po}\to x^{\mu}_{\rm eq}$ as vectors at $z$, namely
\begin{equation}\label{C-60}
R_{\theta}= -R_y, \quad\quad R_{\varphi}=R_x,
\end{equation}
where $R_{\alpha}$ stands for any of the RP (obviously, $R_t$ and $R_r$
do not change). Note also the relations
$u_{\theta}=-u_{y}(z)(=0)$ and $u_{\varphi}=u_{x}(z)$
($u_t$ and $u_r$ do not change).


\end{document}